\documentstyle[prd,aps,floats,psfig]{revtex}
\draft
\begin{document}
\renewcommand{\topmargin}{0.0cm}
\renewcommand{\textheight}{20cm}
\newcommand{\rms}{\rm\scriptstyle}
\newcommand{\rmss}{\rm\scriptscriptstyle}
\newcommand{\nub}{\overline{\nu}}
%

\title{Precise Measurement of Neutrino and Anti-neutrino Differential Cross
Sections}

\author{M. Tzanov, D. Naples, S. Boyd, J. McDonald, V. Radescu} 
\address{Department of Physics, University of Pittsburgh, PA 15260}
\author{R.~A.~Johnson, N.~Suwonjandee, M.~Vakili}
\address{University of Cincinnati, Cincinnati, OH 45221}
\author{J.~Conrad, B.~T. Fleming, J.~Formaggio, J.~H.~Kim, S.~Koutsoliotas,\\
C.~McNulty, A.~Romosan,  M.~H.~Shaevitz, P.~Spentzouris,\\ E.~G.~Stern,
A.~Vaitaitis, E.~D.~Zimmerman}
\address{Columbia University, New York, NY 10027}
\author{R.~H.~Bernstein, L.~Bugel, M.~J.~Lamm,
W.~Marsh, P.~Nienaber, N.~Tobien, J.~Yu.}
\address{Fermi National Accelerator Laboratory, Batavia, IL 60510}
\author{T. Adams, A.~Alton, T.~Bolton, J.~Goldman, M.~Goncharov}
\address{Kansas State University, Manhattan, KS 66506}
\author{L.~de~Barbaro, D.~Buchholz, H.~Schellman, G.~P.~Zeller}
\address{Northwestern University, Evanston, IL 60208}
\author{J.~Brau, R.~B.~Drucker, R.~Frey, D.~Mason}
\address{University of Oregon, Eugene, OR 97403}
\author{S.~Avvakumov, P.~de~Barbaro, A.~Bodek, H.~Budd,
D.~A.~Harris, \\K.~S.~McFarland, W.~K.~Sakumoto,U.~K.~Yang}
\address{University of Rochester, Rochester, NY. 14627}

\maketitle

\date{\today}

\begin{abstract}
The NuTeV experiment at Fermilab has obtained a unique
high statistics sample of neutrino and anti-neutrino
interactions using its high-energy sign-selected
beam. We present a measurement of the differential cross 
section for charged-current neutrino and anti-neutrino 
scattering from iron.
Structure functions, $F_2(x,Q^2)$ and $xF_3(x,Q^2)$,
are determined by fitting the inelasticity, $y$, dependence of the 
cross sections.
This measurement has significantly improved systematic
precision as a consequence of more precise understanding 
of hadron and muon energy scales.
\end{abstract}
\pacs{PACS numbers: 12.38.Qk, 13.15.+g, 13.60.Hb }

\section{Introduction}
\label{sec:intro} 

Deep inelastic scattering (DIS), the scattering of a high energy
lepton off a quark inside a nucleon, has been a proving ground
for QCD, the theory of strong interactions. Charged-leptons
and neutrinos
have been used to measure parton densities and their QCD evolution
with high-precision over a wide range in $Q^2$.
Uniquely, neutrino DIS, via the weak interaction probe, allows 
simultaneous measurement of two structure functions:
$F_2(x,Q^2)$ and the parity-violating structure function, 
$xF_3(x,Q^2)$.

In this paper we present a new measurement of high-energy neutrino and 
anti-neutrino differential cross sections from high-statistics
data samples. The differential dependence of neutrino and 
anti-neutrino cross sections
on Bjorken scaling variable, $x$, and 
inelasticity, $y$, provide the most model-independent physics observable
for this process \cite{barone}.
Previous high-statistics measurements of the 
neutrino and anti-neutrino differential cross section 
have been reported \cite{cdhswxsec}, \cite{ccfrxsec}. 
This experiment has two improvements: first, a sign-selected beam
allowed separate neutrino and antineutrino running and, second,
a calibration beam continuously measured the detector's response.
The largest experimental uncertainties in previous measurements 
arose from knowledge of energy scale and detector 
response functions \cite{nurev},
\cite{qcd_hand}. NuTeV
addressed this by using a dedicated calibration beam 
of hadrons, electrons and muons 
that alternated with neutrino running once every minute throughout the 
one year data-taking peroid. The calibration beam was used to 
measure the detector response for hadrons and muons
over a wide range of energies (5-200~GeV).
Detector response functions and energy scales for muons
and hadrons were mapped over the full active area of the detector.
Energy scales for muons and hadrons were 
determined to a precision of 0.43\% for hadrons
and 0.7\% for muons \cite{nim}. 
NuTeV's other improvement was separate neutrino and antineutrino
running. NuTeV ran in two modes, ($\nu$ and $\nub$),
with the muon spectrometer polarity always set to focus the primary
charged-lepton from the interaction vertex 
({\em e.g.} $\mu^-$ in $\nu$-mode or $\mu^+$ in $\nub$-mode).
In determining the charged-current differential cross sections,
this allowed better and more uniform acceptance in the two running
modes and removed ambiguity in the muon sign determination 
present in a mixed $\nu$ and $\nub$ beam.

The rest of this paper is organized into four parts;
Section II describes the NuTeV detector, 
Section III gives the cross section extraction method and results, 
and Section IV presents the extracted structure functions
and the results are discussed in Section V.

\section{NuTeV Experiment}

\begin{figure}
\centerline{
\psfig{file=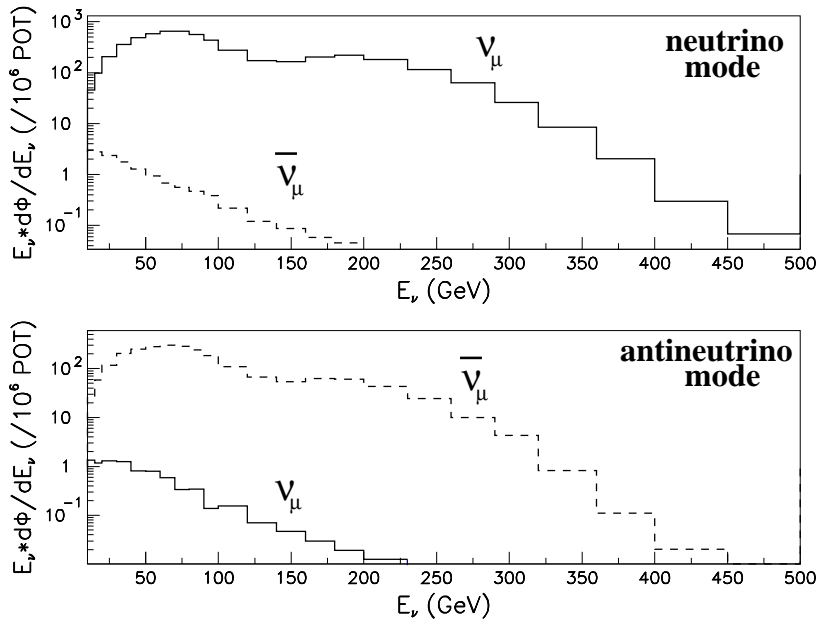,width=0.6\columnwidth}}
\vspace{0.05in}
\caption{Predicted energy spectrum of interacted muon-neutrinos 
anti-neutrinos when the SSQT is set to select neutrinos (top)
and antineutrinos (bottom).}
\label{fig:nuflux}
\end{figure}

The NuTeV experiment collected data during 1996-97  using
separate high-purity $\nu_{\mu}$ and $\nub_{\mu}$ beam produced by the 
Sign-Selected Quadrupole Train (SSQT) beamline. 
A dipole magnet after the one-interaction-length beryllium oxide
production target allowed the sign of secondary particles to be selected.
Neutrinos (or anti-neutrinos) are produced when sign-selected 
secondary pions and kaons decay in the $440$~m decay region
located just downstream of SSQT optics. The NuTeV neutrino detector is 
$1.4$~km downstream of the beryllium oxide production target. 
Neutrino energies ranged from 30-500~GeV.
Figure \ref{fig:nuflux} shows a prediction of the interacting 
$\nu_{\mu}$ and $\nub_{\mu}$ flux in each mode with it's small
contribution from the wrong-sign background.
The interacted neutrino fraction from 
$\overline{\nu}$($\nu$) in $\nu$($\overline{\nu}$)-mode
is $3\times10^{-4}$ ($4\times10^{-3}$).

The NuTeV/CCFR detector consisted of an 18~m long, 690~ton
target calorimeter with a mean density of ${\rm 4.2~
g/cm^3}$, followed by a 420~ton iron toroidal spectrometer. The
target was composed of 168 steel plates, each ${\rm 3m \times 3m \times
5.15cm}$, instrumented with liquid scintillator counters placed
every two steel plates and drift chambers spaced 
every four plates. NuTeV refurbished the CCFR detector by replacing
the scintillator oil and reconditioning the drift chambers. 

Immediately downstream of the target-calorimeter
was a magnetized iron toroidal spectrometer with inner radius 
12.5~cm and outer radius 175~cm used to measure the
momentum of high-energy muons exiting the downstream end of the target. 
The toroid spectrometer consisted of three magnetized sections
each followed by a drift chamber station.
Two additional drift chamber stations were located a few meters
downstream of the last magnetic section to analyze 
the highest momentum muons. 
The magnetic field in each toroidal section ($\sim$15~kG) was 
produced by four copper coils which emerged through the center hole.
The magnetic field is azimuthal
everywhere except for a small radial component in the region 
of the supporting feet and air gap between top and bottom halves
of the washers. The azimuthal component of the field in the first
toroid had an additional small asymmetry (with respect to vertical)
due to a shorted coil on the west side. The detailed geometry of 
the spectrometer, (including the missing coil), was input to 
ANSYS simulation \cite{ansys} to compute the magnetic field map.

\begin{figure}
\centerline{
\psfig{file=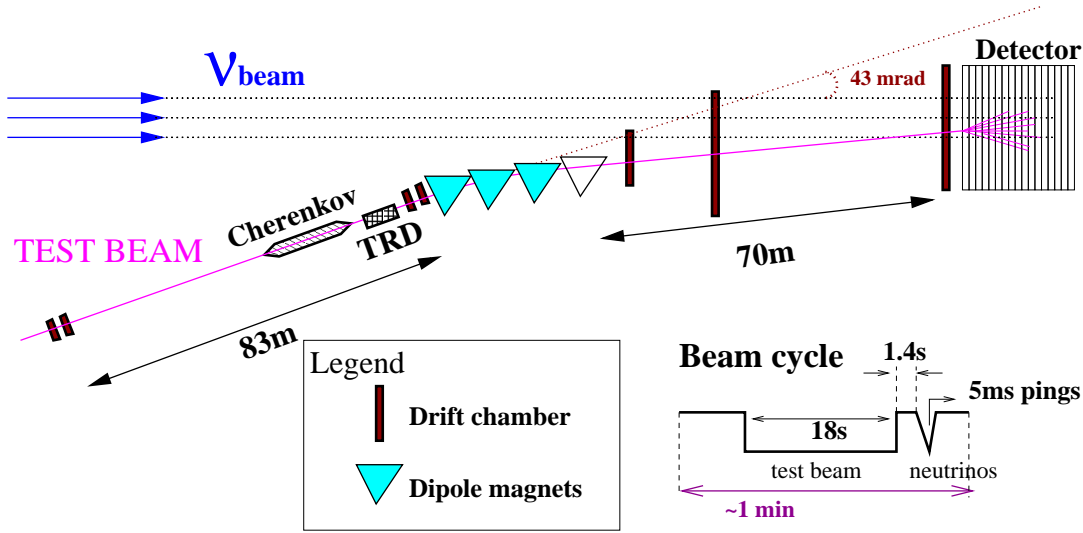,width=0.8\linewidth
}}
\vspace{0.05in}
\caption{Sketch of calibration beam spectrometer configuration. The 
calibration beam ran alternating with neutrinos once a minute 
for the entire NuTeV data-taking period.}
\label{fig:tb}
\end{figure}

A dedicated {\em in situ} calibration 
beam was used to determine the energy response of the calorimeter and
spectrometer to hadrons, muons, and electrons and to map the 
response over the face of the detector. The tolerance of the
calibration spectrometer to measure a beam particle's
absolute momentum was 0.3\%. 
The configuration is shown in Figure \ref{fig:tb}.
Details of the calibration 
and response measurements are given in \cite{nim}. 
The hadronic resolution of the calorimeter was determined to be
$\frac{\sigma}{E}=\frac{86\%}{\sqrt E} \oplus 2.2\%$ with 
an absolute scale uncertainty of $\frac{\delta E}{E}=0.43\%$. The
latter uncertainty is dominated
by the statistical precision in calibrating the 
time dependence of the counter response over
a transverse grid for each counter.

The absolute scale of the toroid spectrometer
was calibrated with muons
that were steered over the active area of the spectrometer.
The magnetic field map from the ANSYS simulation 
was checked with the 50 GeV muon calibration beam. 
The mean reconstructed muon momentum compared with
the muon calibration beam had a 
1$\sigma$ variation of 0.63\% over the active area of the toroid.
This field map uncertainty is attributed to variation in magnetic 
susceptibility and thickness of the steel. The field map determination
dominates the absolute muon energy scale uncertainty of 0.7\%. 

\section{Cross Section Measurement}

In charged-current (CC) neutrino DIS the $\nu$ scatters off a quark
in the nucleon via exchange of a virtual $W$-boson. The cross section
can be expressed in terms of
structure functions, $2xF_1(x,Q^2)$, $F_2(x,Q^2)$, and $xF_3(x,Q^2)$\\
\begin{eqnarray}
\label{eqn:disxsec}
\frac{d^2\sigma^{\nu(\overline{\nu})}}{dxdy} &=& \frac{G_F^2 M
 E}{\pi} {\Big(}\Big[1-y(1+\frac{Mx}{2E})
+\frac{y^2}{2}\frac{1+(\frac{2Mx}{Q})^2}{1+R_L}\Big]\\
&& F_2(x,Q^2) 
\pm \Big[y-\frac{y^2}{2}\Big]xF_3(x,Q^2){\Big)}
\nonumber
\end{eqnarray}
where $G_F$ is the Fermi weak coupling constant, $M$ is the
proton mass, $E_\nu$ is the incident neutrino energy in the lab frame,
and $y$, the inelasticity, is the fraction of energy transferred
to the hadronic system. The $xF_3$ term is added for neutrino
interactions and is subtracted for antineutrinos.
$R_L(x,Q^2)$, the ratio of the cross section for scattering
from longitudinally to transversely polarized W-bosons 
relates $F_2(x,Q^2)$ and $2xF_1(x,Q^2)$
$$ 2xF_1= F_2\Big(\frac{1+{(\frac{2Mx}{Q})^2}}{1 + R_L(x,Q^2)}\Big).$$
Structure functions depend on $x$, the Bjorken scaling variable,
and $Q^2$, the four momentum squared of the virtual W-boson.

Relativistic invariant kinematic variables, $x$, $y$, and $Q^2$ can
be evaluated in the lab frame using the three experimentally
measured quantities: $E_\mu$, energy of the
outgoing primary charged-lepton, $E_{\rms HAD}$ the energy deposited
at the hadronic vertex, and $\theta_\mu$, the scattering angle of
the primary muon.
$$x=\frac{4E_\nu E_\mu sin^2\frac{\theta_\mu}{2} }{2ME_{\rms HAD}},$$
$$y=E_{\rms HAD}/E_\nu,~~\rm{and}$$
$$Q^2=2MxyE_\nu$$
where the neutrino energy, $E_\nu=E_\mu+E_{\rms HAD}$, is reconstructed
from the measured final state particle energies.

\subsection{Event Reconstruction}

Events used in this analysis were triggered by the presence
of a penetrating muon track determined by in-time
hits in scintillation counters in the most downstream region
of the target calorimeter and in the first station of the
toroid spectrometer. This allowed acceptance for charged
current events down to zero hadron energy.

$E_{\rms HAD}$ is determined by summing the pulse heights of
consecutive counters from
just downstream of the event vertex to five counters beyond the
end of the shower region. Longitudinal position of the
event vertex is defined as the first of at least two consecutive
counters with greater than four times the energy deposited
by a minimum ionizing particle (MIP).
\footnote{The definition of one MIP used in NuTeV is the following: 
the mean energy deposited by a 77~GeV muon in one counter
determined using a truncated mean procedure see \cite{nim}.}
The end of the shower region is the last counter before three
consecutive counters with less than four MIP's.
Energy deposited by the primary muon in this region is
removed by subtracting the most probable energy loss
for each counter in the sum.
This energy is included in the reconstructed muon energy.

$E_{\mu}$ at the event vertex is reconstructed in two parts:
energy deposited in the target calorimeter (typically less than 10~GeV)
and remaining energy that is measured in 
the downstream toroid spectrometer.
Energy deposited in the target calorimeter includes the
muon energy within the shower region (discussed above)
and energy deposited beyond the shower region before the
muon exits the calorimeter.
The latter is determined from the
pulse height of energy deposited in each counter by the muon.
For small pulse heights, ($<5$ MIPs), the muon energy loss is
assumed to arise from ionization processes and is
converted to GeV using an $E_{\mu}$ dependent conversion function
which was optimized using GEANT to reproduce both
the most probable value and the width of this
component of the energy loss (see \cite{nim}).
For larger pulse heights ($>5$ MIPs), the energy is assumed
to have a contribution from catastrophic processes
({\em i.e.}~bremsstrahlung, pair production). Therefore,
the additional amount above 5 MIPs
is converted to GeV using a calibration
constant determined from the
calorimeter's electromagnetic response.
The contribution to the muon energy resolution from determination of
energy loss in the target is small
for high energy muons and
dominated by a long tail produced by catastrophic energy loss
processes.

\begin{figure}
\centerline{
\psfig{figure=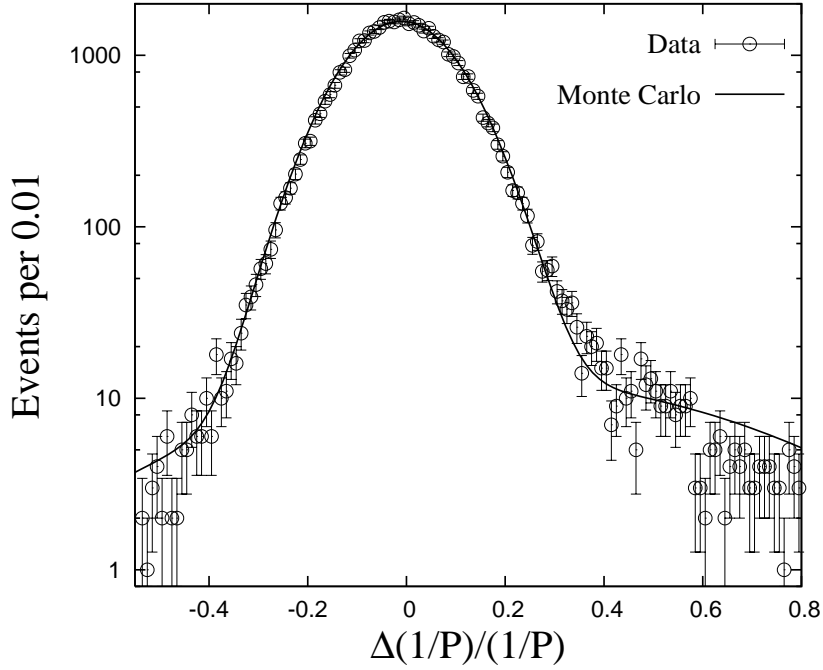,width=0.6\columnwidth
}}
\vspace{0.09in}
\caption{Test beam data (points) with muon momentum
of 100~GeV
measured in the toroid spectrometer compared with double Gaussian
fit parameterization curve.}
\label{fig:pres}
\end{figure}

Muons that enter the toroid spectrometer are focused and
tracked through the spectrometer where they
are momentum analyzed.
Figure \ref{fig:pres} shows the momentum resolution
for the toroid spectrometer determined using test beam muons.
The Gaussian contribution
is dominated by multiple Coulomb scattering (MCS)
and is independent of momentum ($\sim$~11\%).
The high-end tail is due to catastrophic energy loss processes.
Test beam data are used to parameterize the resolution functions
using a fit of the form
\begin{eqnarray}
\frac{\Delta\left(1/P_{\mu}\right)}{\left(1/P_{\mu}\right)}&=&
 exp\left[ -\frac{1}{2}
 \frac{x^2}{\sigma^2_{lead}(P_{\mu})}\right]
 \nonumber
\\
& +& 
R_{tail}(P_{\mu}) exp\left[ -\frac{1}{2}
 \frac{(x - x_{tail}(P_{\mu}))^2}{\sigma^2_{tail}(P_{\mu})}\right],
 \label{musme}
\end{eqnarray}
where $\sigma_{lead}(P_{\mu})$ is 
the width of the leading
Gaussian contribution due to MCS, and $R_{tail}(P_{\mu})$, $\sigma_{tail}(P_{\mu})$
and $x_{tail}(P_{\mu})$ are the normalization coefficient, the width and the offset
of the asymmetric tail of the
resolution function. These parameters are found to have small
energy dependences. The width of the leading Gaussian is parameterized
with the following function
\begin{eqnarray}
\sigma_{lead}(P_{\mu})=\sqrt{A^2 + (BP_{\mu})^2},
 \label{leadsig}
\end{eqnarray}
where $A=0.10$ and $B=4.2\times 10^{-3}$. 
$R_{tail}(P_{\mu})$, 
$\sigma_{tail}(P_{\mu})$, and $x_{tail}(P_{\mu})$ are parameterized
with linear functions.
For a muon entering the toroid with momentum $P_{\mu}=$100~GeV
the resolution function parameters have the following values
$\sigma_{lead}(100) = 0.11$,
$\sigma_{tail}(100) = 0.47$,
$x_{tail}(100) = 0.18$,
and $R_{tail}(100) = 0.079$.

Muon angle is determined using the track vector
in the target calorimeter extrapolated back to the event vertex.
Resolution on muon angle is dominated by multiple scattering in the target
and is determined from GEANT hit-level simulation of the detector.
Angular resolution (in $mRad$) is parameterized as
$$\Delta \theta_\mu= (96.8+0.87~L + 0.24~E_{\rms HAD})/P_{\mu} $$
for a muon with momentum
${P_\mu}$ (in GeV) where $L$ is the track length in the target in
units of counters, and $E_{\rms HAD}$ is the shower energy in GeV.
The small dependence on hadron energy comes from an
$E_{\rms HAD}$ dependent cut excluding tracking chambers near
the event vertex due to the presence of additional
hits in the shower region.

\subsection{Data Selection}

The following criteria were used to select the events for the
cross section sample:

\begin{itemize}
\item [{\em (1)}] {\em Event containment:}
transverse vertex within 125~cm from detector center and
longitudinal vertex
at least four counters beyond the upstream end of the detector
and beginning at least twenty counters from the downstream end.
\item [{\em (2)}] {\em Reconstructed muon energy:}
A single muon in the event with minimum energy,
$E_{\mu}>15$~GeV.
\item [{\em (3)}] {\em Reconstructed muon track in toroid:}
transverse position
within, $15<$ radius $<160$~cm upon entering;
minimum penetration of track to the second chamber station;
and minimum fraction of integrated track in steel $\geq80\%$.
\item [{\em (4)}] {\em Reconstructed hadronic energy:} $E_{HAD}>10$~GeV.
\item [{\em (5)}] {\em Reconstructed neutrino energy:}
$30<E_{\nu}<360$~GeV, is required to
ensure that the flux is well understood and the muon
momentum well reconstructed.
\item [{\em (6)}] {\em Reconstructed $Q^2$:}
$Q^2>1$~GeV$^2$, is required to ensure that non-perturbative
contributions in our cross section model are small.
\item [{\em (7)}] {\em Reconstructed Bjorken-$x$:}
$x<0.8$. Data at higher $x$ are excluded because smearing effects 
are large and our model is not well constrained in this region.
\end{itemize}
The final samples passing these cuts contained 
$8.6\times10^5$ neutrino ($\nu_{\mu}$) and $2.4\times10^5$ 
anti-neutrino ($\overline{\nu}_{\mu}$) events. 

The differential cross section is determined from
the differential number of events and the flux, $\Phi(E)$, at a given
neutrino energy,
\begin{equation}
  \frac{d^2\sigma^{\nu(\overline{\nu})}_{ijk}}{dx dy} \propto
  \frac{1}{\Phi(E_i)}\frac{\Delta N^{\nu(\overline{\nu})}_{ijk}}{\Delta
x_j
    \Delta y_k}.
\label{txs3}
\end{equation}
The quantity $\frac{d^2\sigma^{\nu(\overline{\nu})}_{ijk}}{dx
  dy}$ represents the average differential cross section in bin 
$ijk$.
The absolute flux was not measured in NuTeV. The 
flux was normalized
so that the average NuTeV total cross section from 30-200~GeV 
is equal to the world average value (see section \ref{sec:flux}).

The number of events in a given bin, $N^{\nu(\overline{\nu})}_{ijk}$,
must be corrected for 
bin acceptance due to detector geometry and kinematic cuts,
and for bin migration caused by experimental resolution.
The cross section measured in Equation \ref{txs3} is
corrected to the bin-center value
\begin{eqnarray}
  \frac{d^2\sigma^{\nu(\overline{\nu})}
}{dx dy} (E^c_i,x^c_j,y^c_k)=
  \frac{d^2\sigma^{\nu(\overline{\nu})}
_{ijk}}{dx dy} \times
\frac{S(E^c_i,x^c_j,y^c_k)}{\overline{S}_{ijk}}
\nonumber
\end{eqnarray}
where $S(E^c_i,x^c_j,y^c_k)$ is the differential cross section
evaluated at the bin-center values $E^c_i,x^c_j,y^c_k$
and $\overline{S}_{ijk}$
is the average value of the cross section determined from the
integral over the bin
$$\overline{S}_{ijk}=
\frac{1}{\Delta x_j \Delta y_k}
\int^{x_j+1}_{x_{j}}
\int^{y_k+1}_{y_{k}} \frac{d^2\sigma(E_i,x,y)}{dx dy}~dx~dy.$$
This correction is calculated by integration of the Monte Carlo
model (described in section \ref{sec:mc}) and
is most important at low and high $x$.

\subsection{Neutrino Flux}
\label{sec:flux}

The neutrino (and antineutrino) relative flux as a function 
of energy is determined using the ``fixed $\nu_o$'' method
\cite{nurev}.
Integrating the differential cross section given in eq.~\ref{eqn:disxsec}
over $x$ gives
\begin{eqnarray}
 {\frac{d\sigma}{d\nu}}= A\left( 1 + \frac{B}{A}
  \frac{\nu}{E}
 - \frac{C}{A} \frac{\nu^2}{2E^2} \right)
\label{eqn:dsdnu}
\end{eqnarray} 
where $\nu=E_{\rms HAD}$ and $E$ is the incident neutrino energy. 
The coefficients are given by,

$$A = \frac{G_FM}{\pi}\displaystyle\int F_2(x) dx$$
$$B = -\frac{G_FM}{\pi}\displaystyle\int \Big( F_2(x)  \mp xF_3(x)\Big) dx$$
\begin{eqnarray}
\nonumber
C &=& B - \frac{G_FM}{\pi}\displaystyle\int F_2(x) R_{\rms TERM} dx
\nonumber
\end{eqnarray}
where, $R_{\rms TERM} =
\left(\frac{1+\frac{2Mx}{\nu}}{1+R_{L}}-\frac{Mx}{\nu}-1\right)$,
depends on the longitudinal structure function, $R_L(x)$. 
Multiplying both sides of eq.~\ref{eqn:dsdnu} by the flux  
$\Phi(E)$ gives the number of events
$$      
{\frac{dN}{d\nu}}= \Phi(E) A\left( 1 + 
\frac{B}{A} \frac{\nu}{E} - \frac{C}{A} \frac{\nu^2}{2E^2}
\right).$$
As $\nu\rightarrow 0$ the cross section (eq.~\ref{eqn:dsdnu}) is independent
of energy and therefore the number of events at low $\nu$
is proportional to the flux, $\frac{dN}{d\nu}\rightarrow\Phi(E)A$. 
To minimize the statistical uncertainty, 
data up to $\nu=20GeV$ are included in our flux
sample. Therefore a correction is applied to 
account for the energy dependence as discussed below.
Substituting for the coefficient C, the
relative flux is then given by
$$\Phi({\rm E})=\displaystyle\int _0^{\nu_0} 
\frac{\frac{dN({\rms E})}{d\nu}} { 
      1 + \langle\frac{B}{A}\rangle \left(\frac{\nu}{E} -
      \frac{\nu^2}{2E^2}\right) + \frac{\nu^2}{2E^2}
      \frac{\int F_2(x) R_{\rmss{TERM}} }
       {\int F_2(x)}}\, d\nu.$$
The term $\int F_2(x) R_{\rmss{TERM}} / \int F_2(x)$ is obtained by
integrating the structure functions, calculated using our
Monte Carlo model. 

In reference~\cite{nurev}
it was assumed that the coeficients $A$ and $B$ do not depend
on $\nu$. The integration over $x$ at fixed $\nu$ 
gives an implicit $Q^2$ dependence, $Q^2=2Mx\nu$.
For different values of $\nu$
the integral will be over different ranges in $Q^2$. 
A $\nu$-dependent scaling violations correction is applied
to account for this effect. The correction is obtained by integrating
the structure functions  $F_{2}$ and $xF_{3}$ over $x$ at
fixed $\nu$, using our Monte Carlo model. As described below,
this correction shifts the measured value of $\frac{B}{A}$ and has a 
small effect on the extracted flux.

The flux data sample consists of events which are contained in the
detector, have a well constructed muon with
minimum energy $E_{\mu}>15$~GeV, neutrino energy in the range 
$30<E<360$~GeV, and $E_{\rms HAD}<20$~GeV.
The $E_{\rms HAD}$ cut makes this sample orthogonal to the sample
used to measure the cross sections. 
The data are corrected for acceptance and detector effects using
our Monte Carlo model. Corrections were also applied to remove QED radiative 
effects using~\cite{bardin} and for the charm production threshold
using a leading-order slow rescaling 
model with the charm mass parameter, $m_c=1.40\pm0.18$ 
(see Appendix).
Radiative corrections range from -2\% at 30~GeV to 4\% at 290~GeV.
The charm production correction decreases with energy and is about 5\% 
for the flux sample at 30~GeV.

The coefficient $\frac{B}{A}$ is
determined from a fit to $\frac{dN}{d\nu}$ over the range $5<\nu<20$~GeV.
Limiting $\nu$ to above 5~GeV reduces the contribution
from quasi-elastic and resonance processes, which are difficult to
model. 
The coefficients of the fit 
parameters $A$ and $B$ are modified to remove the $\nu$ dependence due to
scaling violations.
The fit is performed for each energy and the average
value of $\langle\frac{B}{A}\rangle$ over all energy bins is used.
We obtain $\langle\frac{B}{A}\rangle=
-0.34 \pm 0.04$ for neutrinos and $-1.68 \pm 0.03$ for anti-neutrinos. 
The effect of scaling violations is to shift
the average value of $\langle\frac{B}{A}\rangle$ by $+$0.13 
for neutrinos and $+$0.03 for anti-neutrinos. 
The effect on the extracted neutrino flux ranges from 3.5\% at 35~GeV
and is negligible above 120~GeV for neutrinos. For anti-neutrinos
the correction to $\langle\frac{B}{A}\rangle$ is small
and the effect on the flux is less than 0.8\%
for all energies. Because the scaling violations correction is
calculated from a model we estimate a systematic uncertainty in 
$\langle\frac{B}{A}\rangle$ to be 0.05(0.04) for neutrinos(anti-neutrinos).
This theoretical uncertainty is obtained by using an alternative 
model \cite{htby} to 
evaluate the scaling violations and to extract $\langle\frac{B}{A}\rangle$.
The value of $\frac{B}{A}$ computed directly from the NuTeV cross
section model (at $\nu=20$~Gev) gives -0.29 
for neutrinos and -1.66 for antineutrinos
which compares well with the value computed for the alternative 
cross section model (Bodek-Yang) from reference \cite{htby}
which gives -0.28 and -1.68, respectively. 

The total cross section is used to normalize the flux. A sample of
events are selected which are contained in the detector and 
have well constructed muon with minimum energy $E_{\mu}>15$~GeV.
The total neutrino(antineutrino) cross section is
$$\frac{\sigma^{\nu(\overline\nu)}(E)}{E}=\frac{N^{\nu(\overline{\nu})}(E)}{E\cdot\Phi({\rm E})},$$
where $N^{\nu(\overline{\nu})}(E)$ is the number of total cross 
section events, corrected for acceptance and detector effects, and
$\Phi({\rm E})$ is the relative flux.

\begin{figure}
\centerline{
\psfig{figure=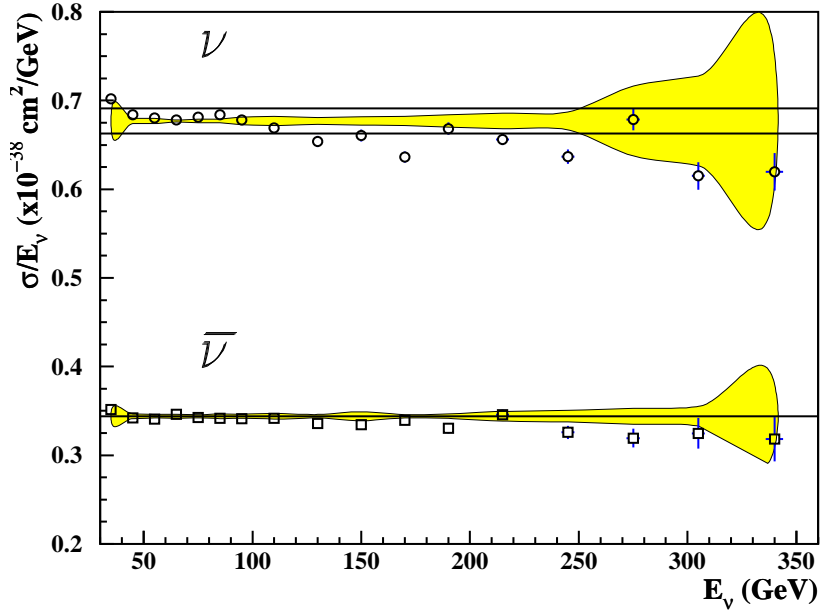,width=0.6\columnwidth
}}
\vspace{0.07in}
\caption{Total cross section, $\frac{\sigma}{E}$, as a function
of energy for neutrino (circles) and anti-neutrino (squares) interactions. The error bars are statistical uncertainty and the yellow band
shows the size of the systematic uncertainty.}
\label{flux}
\end{figure}

The flux is normalized using the world average
neutrino cross section from 30-200~GeV \cite{seligthesis}.
$$\frac{\sigma^{\nu}}{E}=0.677 \pm 0.014 ~~ \times 10^{-38}
~\frac{\rms cm^2}{\rms GeV}$$
The cross section normalization uncertainty, 2.1\%, arises from 
the quoted errors on the world average absolute neutrino cross section.
Figure \ref{flux} shows the energy dependence of the
total cross section, $\frac{\sigma}{E}$.
The total neutrino and antineutrino cross sections are linear with energy
to better than 2\% over our energy range.
The relative $\nub$ to $\nu$ cross section,
$r=\frac{\sigma^{\nub}}{\sigma^{\nu}}$, can be measured from 
NuTeV data alone which gives 
a value of $r=0.505\pm0.0018$(stat)$\pm0.0029$(syst).
This measurement is the most accurate determination to date and 
it agrees with the previous
world average value of $r=0.499\pm0.007$.

\subsection{Cross Section Extraction}
\label{sec:mc}

The Monte Carlo simulation, used to account for acceptance and
resolution effects, requires an input cross section model
which is iteratively tuned to fit our data.
Our cross section model, described in the Appendix, is based on 
a leading order prescription from Buras and Gaemers\cite{bg} 
that is modified to incorporate higher-order corrections such as
$R_L(x,Q^2)$, charm mass, and higher-twist effects which are 
important at low $Q^2$. Cross section data are fit to determine 
empirically a set of parton distribution functions. 
To model regions at the edge of 
our data sensitivity we use additional
input to constrain the cross section. At high-$x$ and 
low $Q^2$ we model higher-twist contributions 
following reference \cite{arie} by incorporating
charged-lepton data in the fit.
At low $x$ and low $Q^2$, (below 1.35~GeV$^2$), where the
Buras-Gaemers parameterization is not well behaved,  
the shape of GRV94LO \cite{grv94lo} is used.

The cross section, flux, and the model parton distribution functions
($PDF$s) are obtained by a reiterative extraction and refitting 
proceedure. An initial set of model parameters from 
CCFR\cite{ccfrxsec} are used to extract an initial flux and cross section.
From these we perform a fit to obtain a set of $PDF$s which are then
used to extract a new flux and cross section.
The proceedure is reiterated until the relative change in 
cross section value from one
iteration to the next averaged over all the data points
is less than 0.1\% (this occurs within three iterations). 
New radiative corrections calculated from \cite{bardin} are 
computed for each new cross section fit. 
After the final iteration, we exclude kinematic bins 
where the number of events generated in that
bin account for less than 20\% of the events. 
This reduces the contribution
from the high $E_{\mu}$ tail where smearing dominates the distribution.

\begin{figure}
\centerline{
\psfig{figure=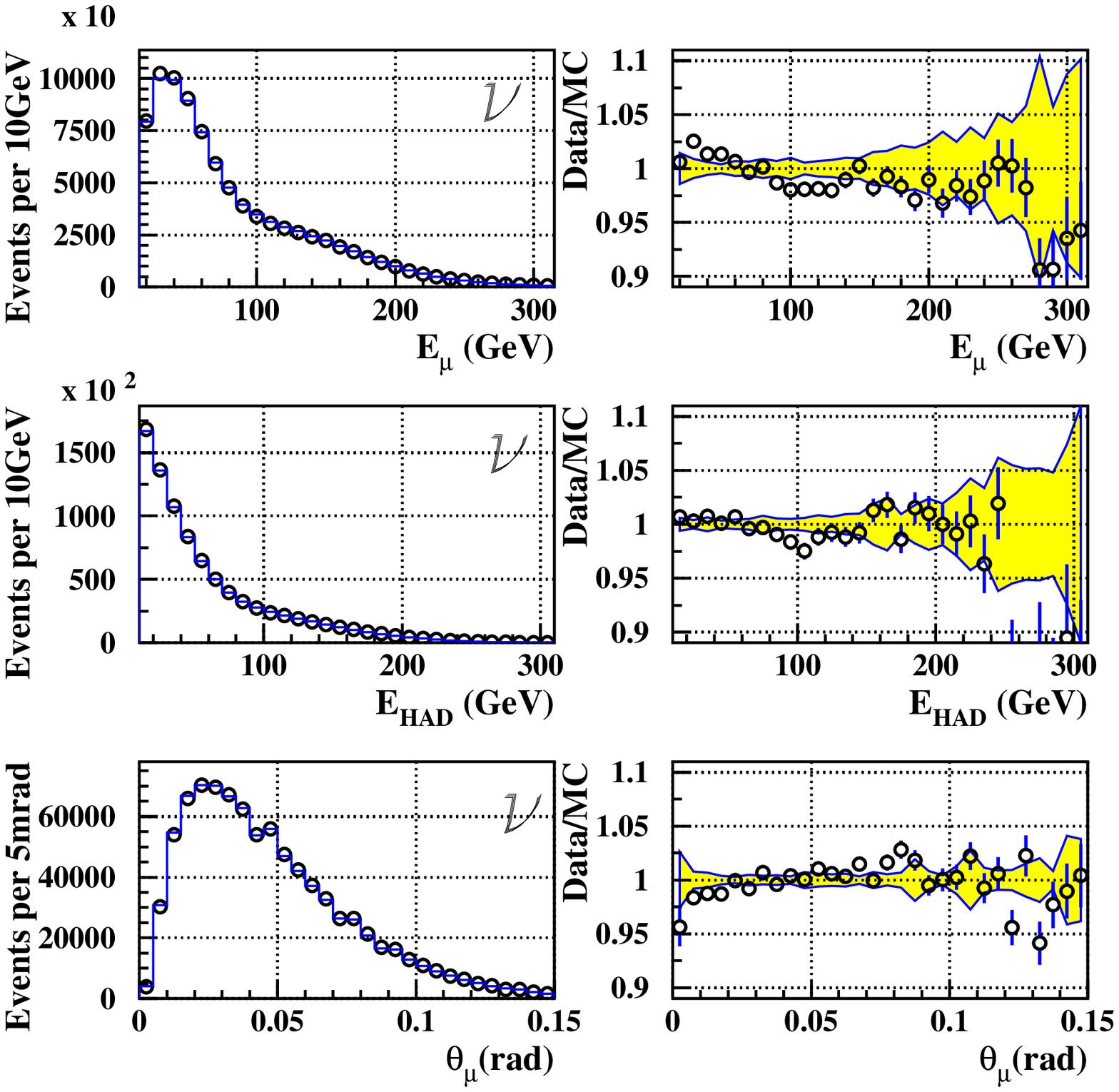,width=0.55\columnwidth}}
\centerline{
\psfig{figure=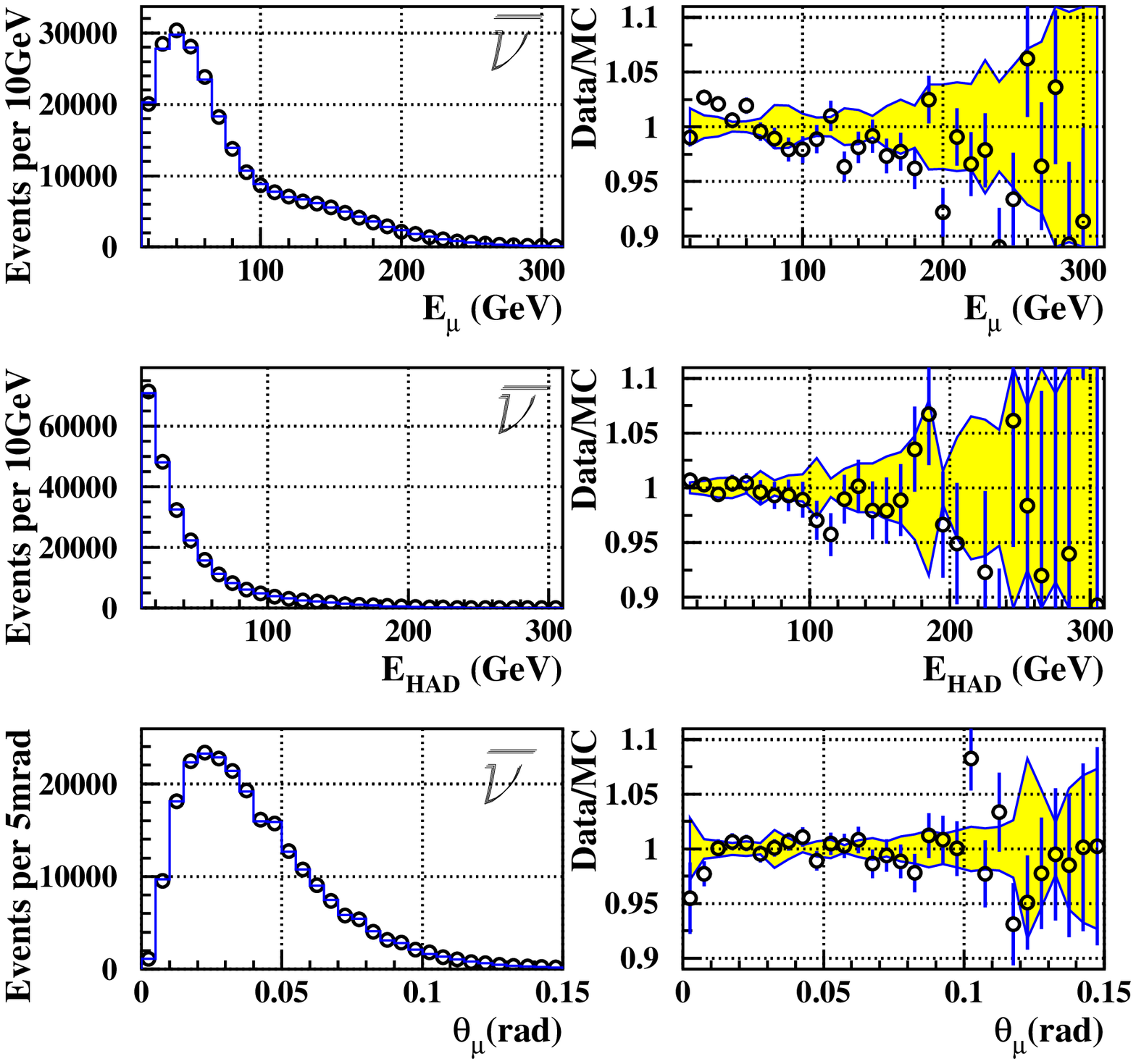,width=0.55\columnwidth}}

\caption{Comparison of data distributions for
kinematic variables, $E_\mu$, $E_{HAD}$, and $\theta_\mu$
with Monte Carlo simulation for neutrinos (top) and anti-neutrinos (bottom). 
The points are data and the curve is the Monte Carlo model. Ratio of data
to Monte Carlo is plotted at the right of each distribution.
The error bars are statistical uncertainty and the yellow band
shows the size of the systematic uncertainty.}
\label{datamc}
\end{figure}

Figure \ref{datamc} shows a comparison of the distributions of the
three kinematic variables measured in data with those 
determined from the Monte Carlo model.
The data versus model $\chi^2$ including systematic uncertainties is
$\chi^2/dof=2225/2599$. 
If full point-to-point data correlations
are included in the $\chi^2$ the fit quality worsens to 
$\chi^2/dof=3534/2599$. (The data correlation matrix 
is discussed in Section \ref{sec:syst}).
The inclusion of the extremely precise charged-lepton
data in our model fit (see Appendix) has the effect of 
systematically pulling the $Q^2$ dependence of the model into
agreement with the charged-lepton data and worsens the 
quality of the model fit with NuTeV data.
Alternative models based on global parton distribution fits
give a significantly poorer $\chi^2$ with our data; the 
Bodek-Yang model from reference \cite{htby} gives
$\chi^2=5969/2599$ and TRVFS(MRST99) \cite{trvfs} gives
$\chi^2= 5000/2599$. 
Model sensitivity in the cross section measurement is small
(see section \ref{sec:syst}).

\subsection{Results}

\begin{figure}
\centerline{
\psfig{figure=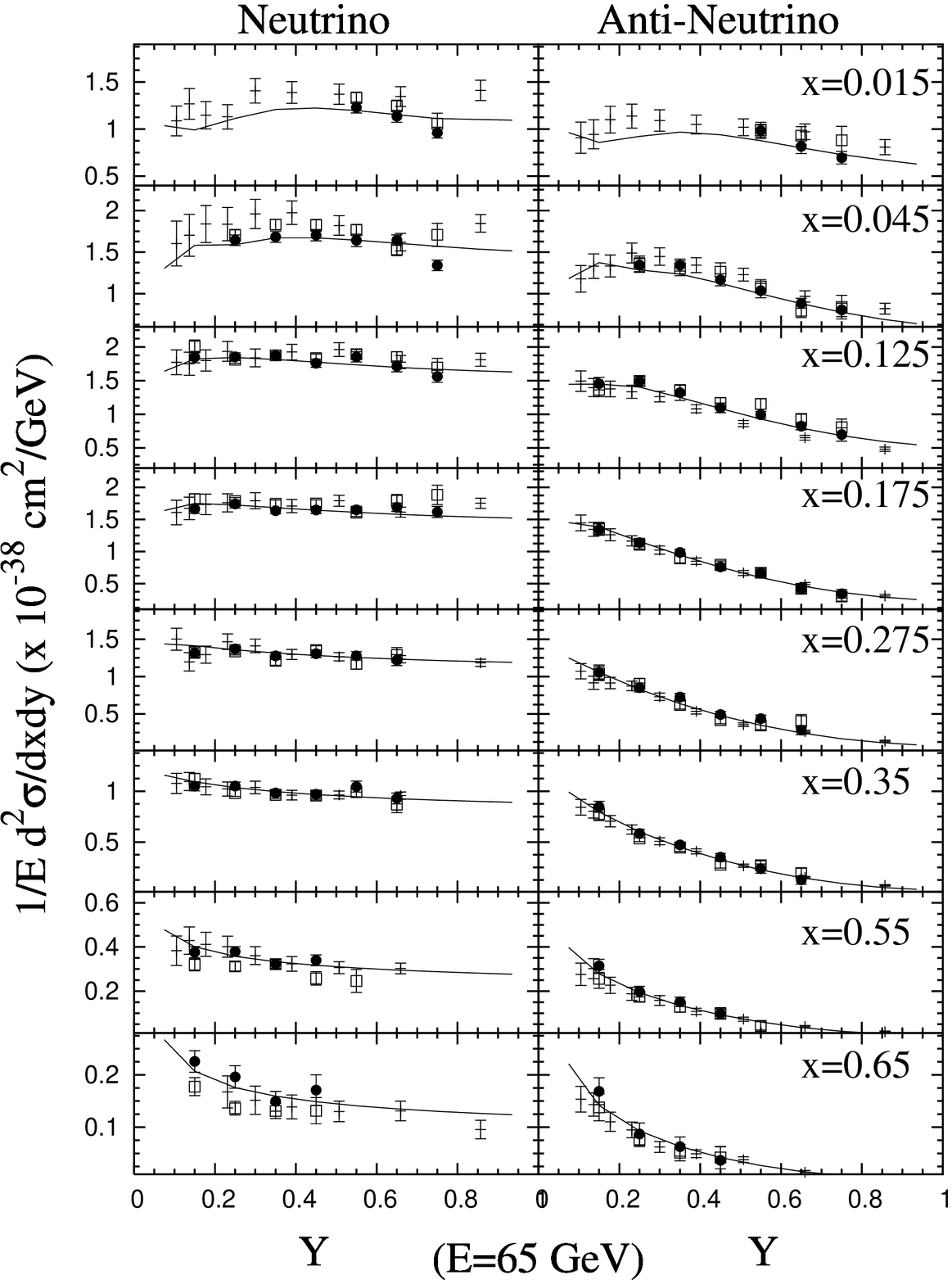,width=0.6\columnwidth
}}
\vspace{0.07in}
\caption{Differential cross sections in $x$ bins for
neutrinos (left) and anti-neutrinos (right) at $E=65$~GeV.
Points are NuTeV (filled circles),
CCFR (open squares), and CDHSW (crosses).  
Error bars show statistical and systematic errors in quadrature.
Solid curve shows fit to NuTeV data.
($x=$0.08, 0.225, 0.45, and 0.75 bins are not shown).}
\label{diffxsec1}
\end{figure}

\begin{figure}
\centerline{
\psfig{figure=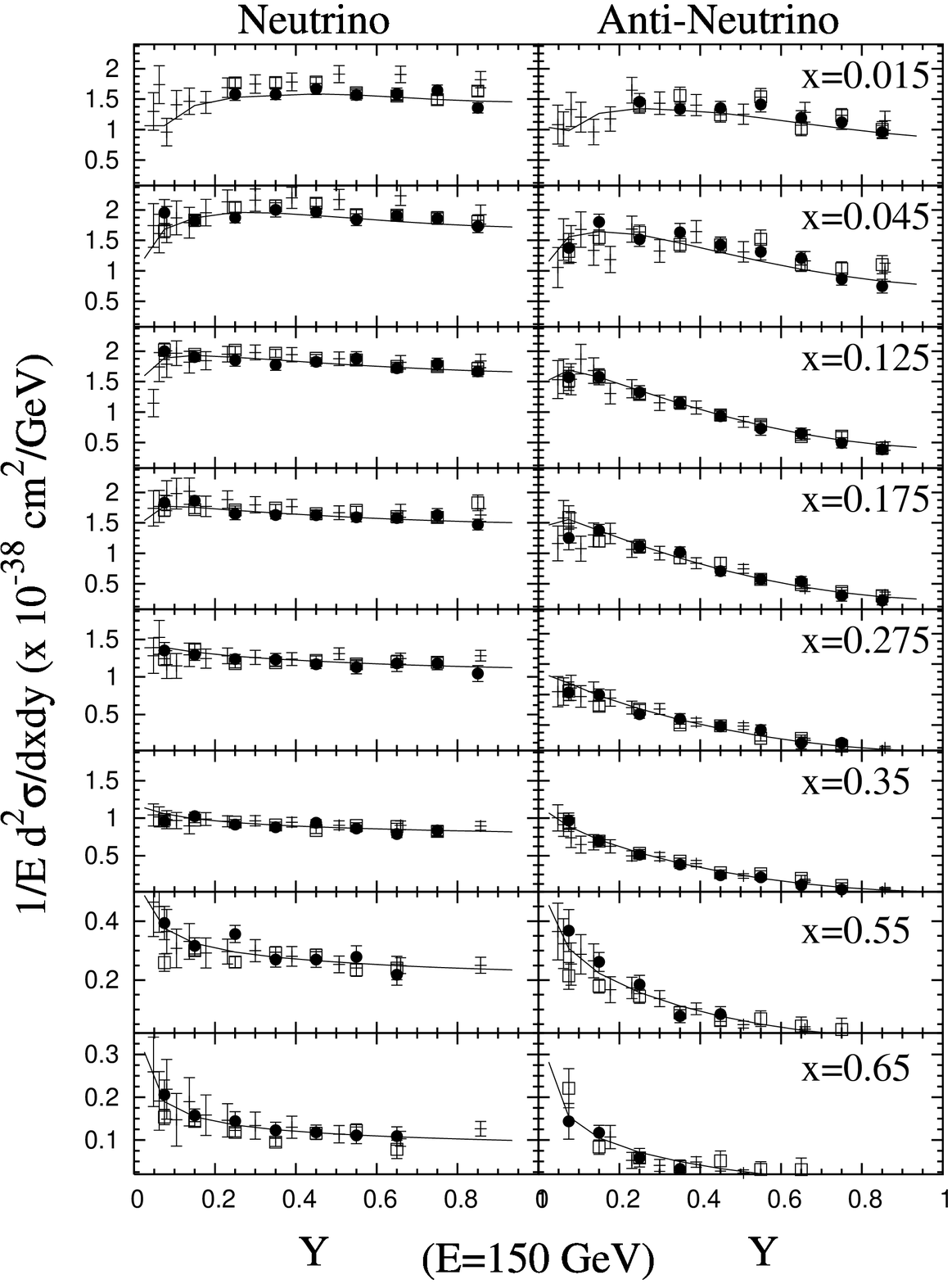,width=0.6\columnwidth
}}
\vspace{0.07in}
\caption{Differential cross sections in $x$ bins for
neutrinos (left) and anti-neutrinos (right) at $E=150$~GeV.
Points are NuTeV (filled circles),
CCFR (open squares), and CDHSW (crosses).  
Error bars show statistical and systematic errors in quadrature.
Solid curve shows fit to NuTeV data.
($x=$0.08, 0.225, 0.45, and 0.75 bins are not shown).}
\label{diffxsec2}
\end{figure}

\begin{figure}
\centerline{
\psfig{figure=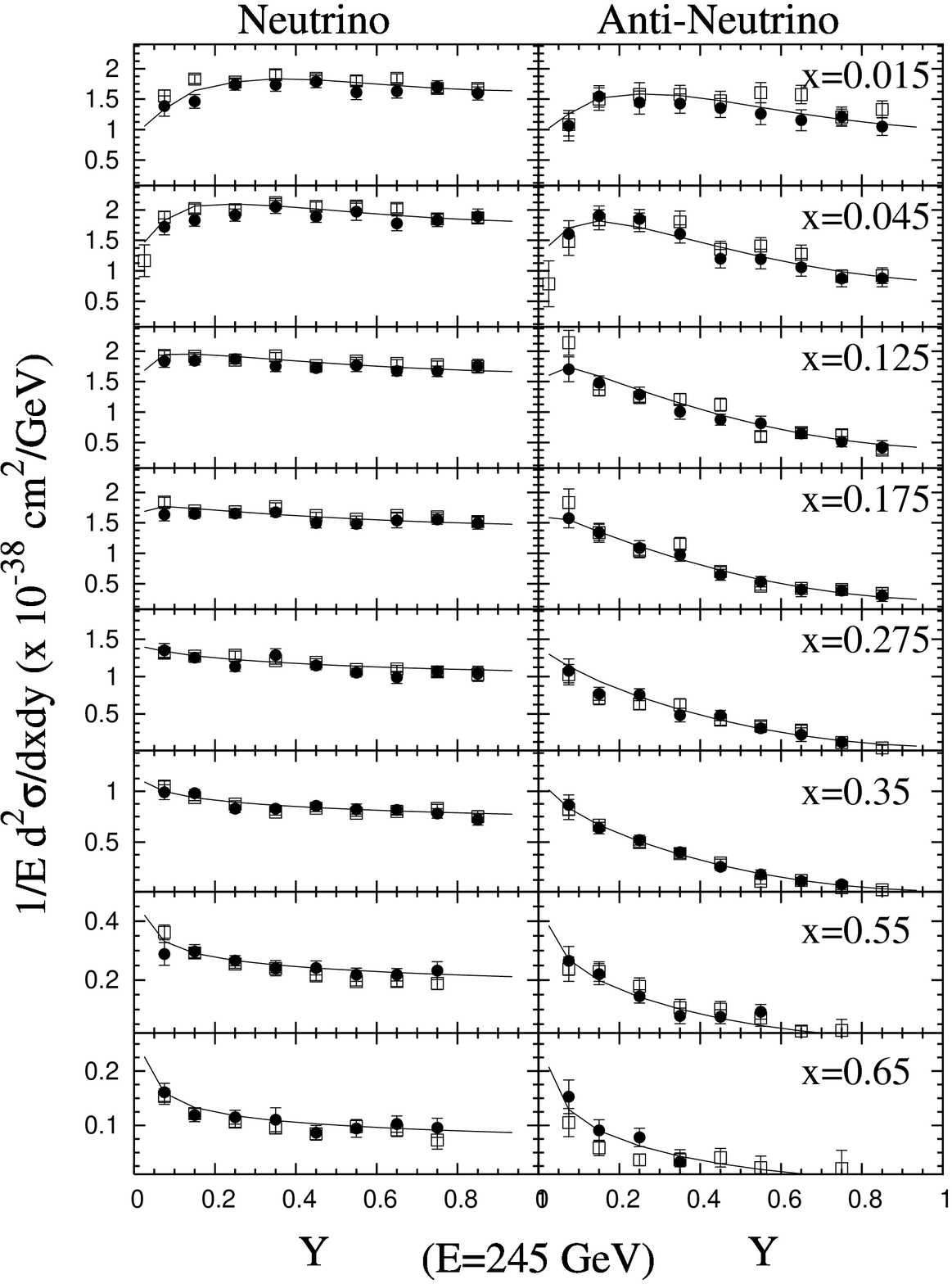,width=0.6\columnwidth
}}
\vspace{0.07in}
\caption{Differential cross sections in $x$ bins for
neutrinos (left) and anti-neutrinos (right) at $E=245$~GeV.
Points are NuTeV (filled circles),
CCFR (open squares), and CDHSW (crosses).  
Error bars show statistical and systematic errors in quadrature.
Solid curve shows fit to NuTeV data.
($x=$0.08, 0.225, 0.45, and 0.75 bins are not shown).}
\label{diffxsec3}
\end{figure}

Figures \ref{diffxsec1}-\ref{diffxsec3} show the extracted
$\nu-Fe$ and $\overline{\nu}-Fe$ 
cross sections plotted as a function of $y$ for a
representative sample of $x$ bins at neutrino energies of 
$65$~GeV, $150$~GeV, and $245$~GeV respectively. 
The NuTeV data are compared with measurements from 
CCFR \cite{ccfrxsec} and at lower energies 
with CDHSW data \cite{cdhswxsec}.
The curve shown is the 
parameterization fit to the NuTeV data.
The three data sets are in reasonable agreement in both level and shape
at low and moderate $x$.
There are differences in NuTeV and CCFR cross sections
at $x> 0.40$ where CCFR's measurement for both 
$\nu$ and $\overline{\nu}$ cross sections 
are consistently below the NuTeV result over the 
entire energy range. The level difference in the cross sections
for these bins at a given $x$ is constant over the full $y$ range of the data.
The difference in the neutrino cross sections are
$4\pm1\%$ at  $x=0.45$, $9\pm2\%$  at  $x=0.55$, 
and increases with $x$ up to 
$18\pm2\%$  at  $x=0.65$, (and similarly for 
antineutrinos).
This discrepancy and its probable source are discussed in 
Section \ref{disc}.

\subsection{Systematic Uncertainties}
\label{sec:syst}

We evaluated the contribution from seven experimental systematic uncertainties
on the cross section measurement error. These include uncertainties in the 
following: muon and hadron energy scales,
muon and hadron energy smearing, the charm
mass value $m_c$ and $\frac{B}{A}$ (both used in 
the flux determination),
and the cross section model which is used 
to perform acceptance corrections. 
Each uncertainty is evaluated separately and then
propagated through the fitting proceedure.
The contribution to the cross section 
uncertainty from each systematic error
(except the muon momentum smearing model which is
described below) is evaluated by
re-extracting the cross section with the value of the 
systematic parameter varied alternately by $\pm 1 \sigma$.
The symmetrized difference in each cross section point
is taken to be the  $1 \sigma$ systematic error due to the
uncertainty in the parameter. 

The largest experimental systematic uncertainties are due to the
muon and hadron energy scale
which for NuTeV are 0.7\% and 0.43\% respectively.
The neutrino and antineutrino fluxes are sensitive to 
the charm mass value $m_c$, used in the charm
production model (see Appendix), and the value of 
$\frac{B}{A}$ used to correct the flux data sample.
The uncertainty in the charm mass parameter
is taken to be $\delta m_c=0.18$, which 
is obtained from the weighted average of 
leading-order experimental measurements \cite{rab},\cite{locharm}.
The values of $\frac{B}{A}$ for neutrino and antineutrino cross 
sections and their uncertainties are obtained from fits to the
NuTeV flux sample data as described in \ref{sec:flux}.
An uncertainty of 2.1\% in the absolute flux determination arises from
the normalization to the world average neutrino cross section.
This is treated separately as an 
overall normalization uncertainty in the cross section.

Detector resolution functions for muon and hadron energy 
reconstruction also contribute to the systematic uncertainty since a
different smearing model
will generate different acceptance corrections from our 
Monte Carlo. 
The hadron energy response was determined from a fit to calibration beam data
\cite{nim}. 
The response function is varied using the one 
sigma error from this fit as described above.
The muon momentum smearing model is given by Equation \ref{musme}.
An estimate of the uncertainty due to the model
is obtained using two different functional forms 
for the parameters of the leading Gaussian: the default model
(Equation \ref{leadsig}) 
and a second model which assumes a linear dependence with energy.
Although both models describe the test beam data reasonably well 
their extrapolations differ at higher energies ($p_{\mu}>200$GeV).
The alternative models are used to re-extract the flux and cross section.
The uncertainty is taken to be the point-to-point difference
between the extractions.
The flux is manifestly independent of the muon angle smearing model
and the cross section is highly insensitive. We, therefore,
neglect this systematic uncertainty.

To estimate the uncertainty from the cross section model we vary the
errors in the model fit parameters (see Appendix) 
by one sigma from their fit values.
The resulting uncertainties on the cross section are smaller than 
the statistical precision on a Monte Carlo sample with approximately
twenty times the data statistics. 
The effect of the model uncertainty on the cross section error
is very small and is neglected. 

The NuTeV data are presented along with a full point-to-point
covariance matrix that provides the correlation coefficient between
any two cross section data points. We have found that these 
correlations are large for neighboring bins. Previous measurements
by \cite{cdhswxsec} and \cite{ccfrxsec}
did not provide such a data correlation matrix.
The covariance matrix, $M_{\alpha\beta}$, is given by
$$ M_{\alpha\beta} =
\sum_i^7 \delta_{i|\alpha} \delta_{i|\beta},$$
where $\delta_{i|\alpha}$ is the $1\sigma$ shift in data point $\alpha$ due
to systematic uncertainty $i$. 
The 2.1\% flux normalization uncertainty can be included in the
covariance matrix by adding a term
$$M^{\prime}_{\alpha\beta}= M_{\alpha\beta}+ 
(0.021)^2\frac{d^2\sigma}{dxdy}_\alpha^{D}\frac{d^2\sigma}{dxdy}_\beta^{D}$$
The statistical uncertainty
is added in quadrature to the diagonal elements of the data covariance matrix.

Separate data vectors and covariance matrices are obtained for the neutrino and antineutrino
cross sections. There are 1423 ($N_{DATA}^\nu=1423$) neutrino and 1195 ($N_{DATA}^{\overline\nu}=1195$) antineutrino data points. A $\chi^2$ with respect to a theoretical model can be calculated using 
\begin{eqnarray}
&\chi^2&= \\ 
&=& \sum_{\alpha,\beta=1}^{N_{DATA}^\nu} \left[ \frac{d^2\sigma}{dxdy}_\alpha^{th} - \frac{d^2\sigma}{dxdy}_\alpha^{D} \right]  ({\bf M}^{-1}_\nu)_{\alpha\beta}  \left[ \frac{d^2\sigma}{dxdy}_\beta^{th} - \frac{d^2\sigma}{dxdy}_\beta^{D} \right] 
\nonumber
\\
&+&  \sum_{\alpha,\beta=1}^{N_{DATA}^{\overline\nu}} \left[ \frac{d^2\sigma}{dxdy}_\alpha^{th} - \frac{d^2\sigma}{dxdy}_\alpha^{D} \right]  ({\bf M}^{-1}_{\overline \nu})_{\alpha\beta}  \left[ \frac{d^2\sigma}{dxdy}_\beta^{th} - \frac{d^2\sigma}{dxdy}_\beta^{D} \right],
\nonumber
\end{eqnarray}
where $\frac{d^2\sigma}{dxdy}_\alpha^{D}$ is the measured differential
cross section and
$\frac{d^2\sigma}{dxdy}_\alpha^{th}$ is the model prediction
for data point $\alpha$.

\section{Structure Functions}
\begin{figure}
\centerline{
\psfig{figure=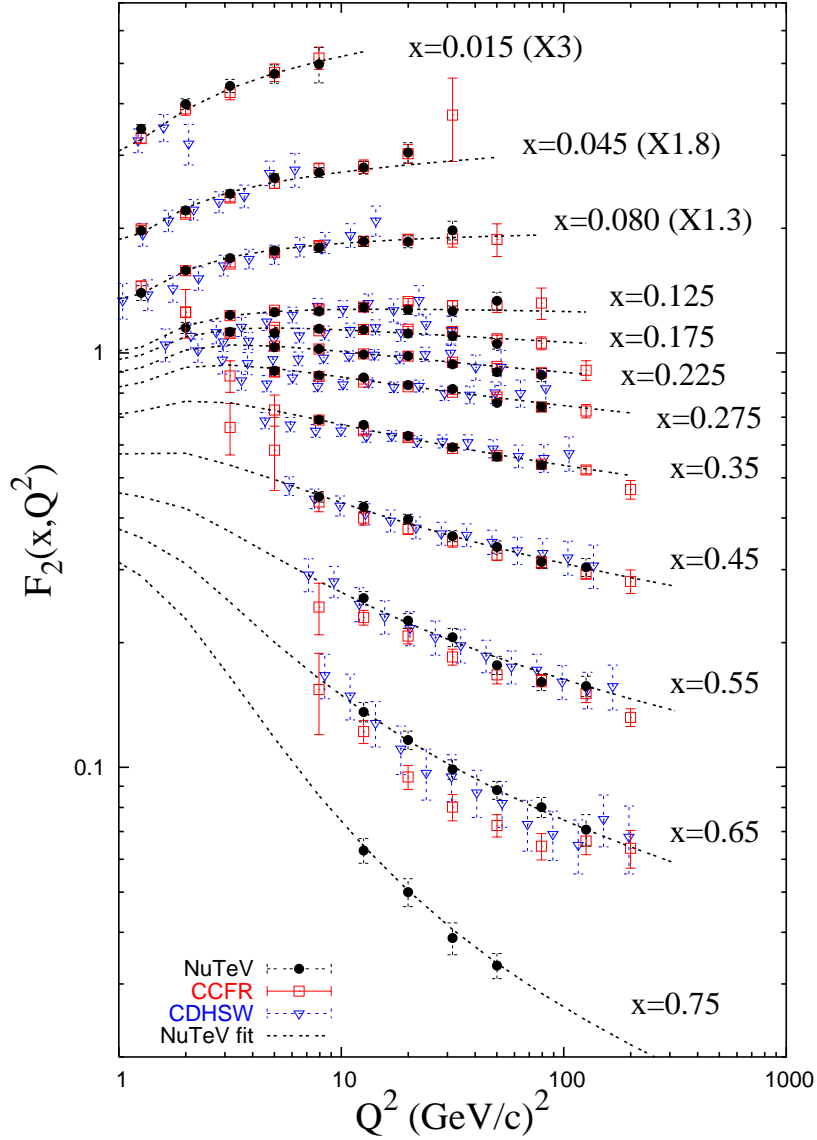,width=0.6\columnwidth
}}
\vspace{0.3cm}
\caption{NuTeV measurement of $F_2(x,Q^2)$ (solid circles)
compared with previous $\nu$-Fe results; CCFR (open circles) and
CDHSW (triangles). The data are corrected to an isoscalar (iron) target
and for QED radiative effects as described in the text.
The curve show the NuTeV model.}
\label{f2}
\end{figure}

Structure functions, $F_2(x,Q^2)$ and $xF_3(x,Q^2)$,
can be determined from fits to linear combinations
of the neutrino and antineutrino differential cross sections.
The sum of the differential cross sections can be expressed as

\begin{eqnarray}
&&{\frac{d^2\sigma}{dx dy}}^{\nu}+{\frac{d^2\sigma}{dx dy}}^{\overline\nu}
={\frac{G_{F}^2
M E}{\pi}}\Big[2\Big(1-y-{\frac{M x y}{2E}} \\
&+& {\frac{y^2}{2}{\frac{1+4 M^2 x^2/ Q^2}
         {1+R_{L}}}}\Big) F_{2} + {y} \Big(1 - {\frac{y}{2}}\Big)
         \Delta xF_{3}\Big].
\nonumber
\end{eqnarray}
where $F_{2}$ is the average of $F_{2}^{\nu}$
and $F_{2}^{\overline {\nu}}$.
The last term is proportional to the difference in $xF_3$ for neutrino
and antineutrino probes,
$\Delta xF_3=xF_3^{\nu}-xF_3^{\overline{\nu}}$, which at leading 
order is $4x\left(s-c\right)$, (assuming 
symmetric $s$ and $c$ seas).
Cross sections are corrected for the excess of neutrons
over protons in the iron target (5.67\%) so that the presented
structure functions are for an isoscalar target. A correction was
also applied to remove QED radiative effects \cite{bardin}.
To extract $F_2(x,Q^2)$ we use
$\Delta xF_3$ from a NLO QCD model as input (TRVFS \cite{trvfs}).
The input value of $R_L(x,Q^2)$ comes from
a fit to the world's measurements \cite{rworld}.
The NuTeV measurement of $F_2(x,Q^2)$ on an isoscalar-iron target
is shown in Figure \ref{f2}.
The structure function,  $F_2(x,Q^2)$ is
compared with previous measurements from CDHSW \cite{cdhswxsec}
and CCFR \cite{ccfrxsec}.

\begin{figure}
\centerline{
\psfig{figure=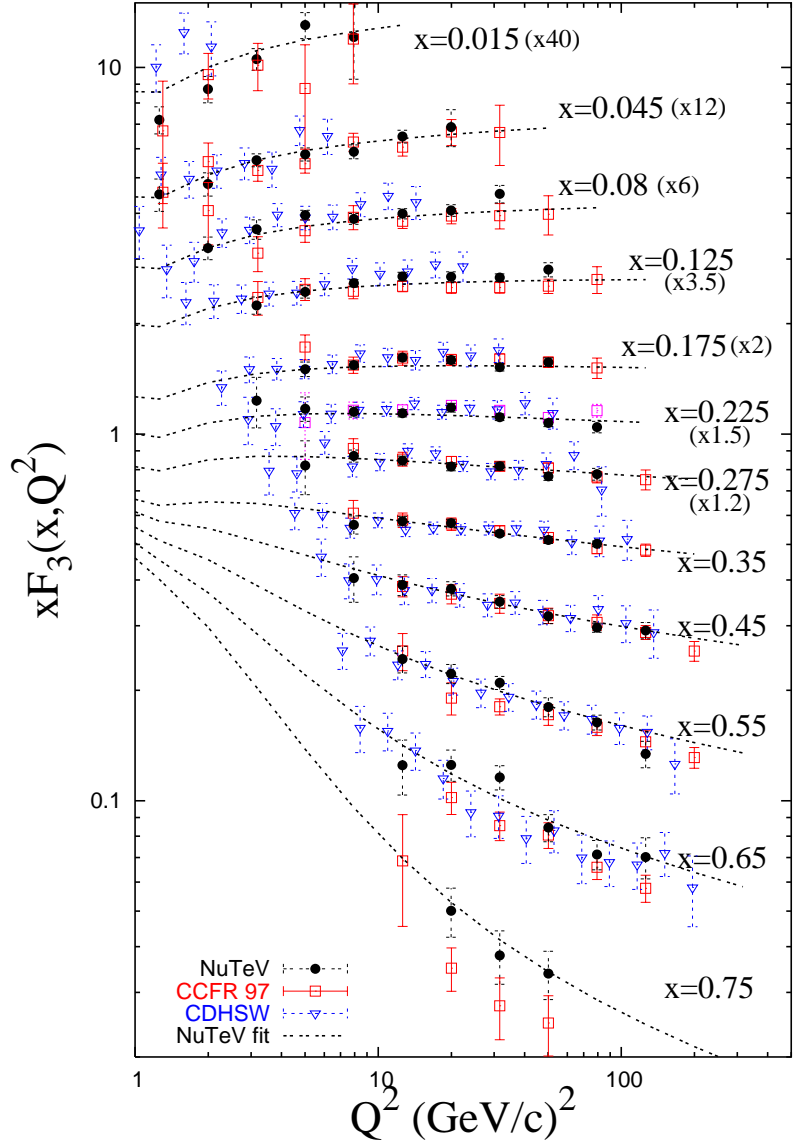,width=0.6\columnwidth
}}
\vspace{0.3cm}
\caption{$xF_3(x,Q^2)$ NuTeV (solid circles)
compared with previous measurements; CCFR97 (open circles) and 
CDHSW (triangles).
The data are corrected to an isoscalar (iron) target
and for QED radiative effects as described in the text.
The curve shows the NuTeV model.}
\label{fig:xf3}
\end{figure}

The difference of neutrino and anti-neutrino differential cross
sections is proportional to the structure function $xF_3(x,Q^2)$,
\begin{equation}
\frac{d^2\sigma^{\nu}}{dxdy} -
      \frac{d^2\sigma^{\overline{\nu}}}{dxdy}=
\frac{2 G_F^2 M E}{\pi}
\left[y-\frac{y^2}{2}\right] xF_3^{\scriptstyle AVG}(x,Q^2)
\end{equation}
where $xF_3^{\scriptstyle AVG}=\frac{1}{2}(xF_3^{\nu}+xF_3^{\overline{\nu}})$,
and the difference between $F^{\nu}_2(x,Q^2)$ and
$F^{\overline{\nu}}_2(x,Q^2)$ is assumed to be negligible.
Figure \ref{fig:xf3}
shows the NuTeV measurement of $xF_3(x,Q^2)$ from fits 
to the cross section difference. 
The structure function,  $xF_3(x,Q^2)$ is
compared with previous measurements from CDHSW \cite{cdhswxsec}
and CCFR97 \cite{seligmanqcd}.

\section{Data Comparisons} 
\label{disc}                                                                           
At moderate $x$, this result agrees well with CCFR over the
full energy and $y$ range of the data, both in level and in shape,
and agrees in level with CDHSW. The CDHSW measurement
has a known $Q^2$ shape difference 
with CCFR \cite{ccfrxsec}, and thus also with this result.
There are differences in NuTeV and CCFR
at $x> 0.40$ where CCFR's measurement for both 
neutrino and antineutrino cross sections 
are consistently below our result for all energies.
This is suprising since NuTeV used the refurbished CCFR detector and
the analysis methods used by the two experiments were very similar.
We discuss below several sources which
may contribute to the high-$x$ cross section difference.

We have determined that the largest single contribution to the
discrepancy is due to a mis-calibration of the magnetic
field map of the toroid in CCFR. 
NuTeV performed thorough calibrations 
of muon and hadron responses in the detector, including mapping
the response over the detector active area and measuring the energy scale
over a wide range of energies \cite{nim}. 
This allowed NuTeV to measure precisely the radial dependence of the magnetic 
field in the toroid. Both experiments used the same muon spectrometer; therefore,
the model of the magnetic field could have a different overall normalization, 
but the radial dependence (which is determined by the geometry of the muon 
spectrometer) should be the same for both. CCFR used one
high statistics muon test beam run aimed at a single
point in the spectrometer to set the absolute energy scale and 
modeled the radial dependence of the magnetic field using 
POISSON \cite{ccfrmag}. 
NuTeV used ANSYS to model the field and compared the
prediction to a precision test beam map data. The width of the residual
fractional difference distribution over the 45 test beam points
is the main contribution to the absolute muon energy scale uncertainty
(0.7\%). If NuTeV uses the CCFR model the result is shifted
to within 1.6 sigma agreement with CCFR at $x=0.65$. This 
accounts for 6\% of the 18\% difference at $x=0.65$.
The field model differences can also be translated
into an effective 0.8\% difference in the muon energy 
scales by integrating the difference in the field models over 
the toroid. 

The cross section model contributes an additional 
$\sim$3\% to the discrepancy seen at x=0.65.
Both experiments determine acceptance corrections using 
an iterated fit to the cross section data. Because the
measured cross sections are different, this necessarily
requires that the respective cross section models 
reflect the data differences.
For example, the NuTeV model is above CCFR by $\sim 20\%$ 
at x=0.65. We have extracted cross sections using CCFR 
cross section model\cite{ccfrxsec} and find that this 
contributes $\sim$3\% to the discrepancy seen at x=0.65.

Other smaller sources for the difference come from 
muon and hadron energy smearing models which 
can also contribute to 
extracted cross section differences through acceptance corrections.
We have determined that using the CCFR muon and hadron smearing models
results in a difference of $\sim 2\%$ at $x=0.65$. 
In NuTeV the hadron energy response
was found to have a small nonlinearity due to the shower $\pi^0$ 
fraction dependence on energy \cite{nim}. This nonlinearity was taken
into account in NuTeV but not in CCFR. The effect of incorporating
the hadron energy scale nonlinearity is small and contributes mainly
to the $Q^2$ dependence. 

All together these three contributions 
account for about two thirds of the high-$x$ cross section difference seen. 
This brings the two measurements within 1.2 sigma agreement 
in the high-$x$ region.

Another significant difference in the two experiments
is that, while both used wide-band beams, NuTeV's beam was
sign-selected. In NuTeV, neutrino and anti-neutrino data were taken
using separate high-purity beams.
This allowed NuTeV to run the detector's toroidal magnetic field
polarity always set to focus
the ``right-sign'' of muon produced in charged-current interactions.
In CCFR, the beam had an 11\% anti-neutrino component and
neutrino and anti-neutrino data were taken simultaneously. 
The toroidal field polarity was reversed perodically 
to alternately focus either $\mu^+$ or $\mu^-$.
To obtain adequate antineutrino statistics CCFR ran approximately 
half the time in focussing mode for each muon sign.
This has two effects on the CCFR analysis that are not 
present in NuTeV. First, the acceptance corrections
are different depending on whether the toroid is set 
to focus or defocus the muon. Second, in defocussing mode 
acceptance corrections are larger and consequently
need to be more accurately determined. 
Acceptance falls off
rapidly with $y$ at high-$x$, especially in defocussing
mode where low-energy and wide-angle muons are deflected outward and
spend less time the toroid magnetic field. 
CCFR had no test beam data for defocussing mode and, while
acceptance corrections were modeled with a field
simulation, the smearing model was assumed to be the same 
in both modes. 
We speculate that this 
difference may also contribute to the discrepancy seen.

\section{Theory Comparisons} 

\begin{figure}
\centerline{
\psfig{figure=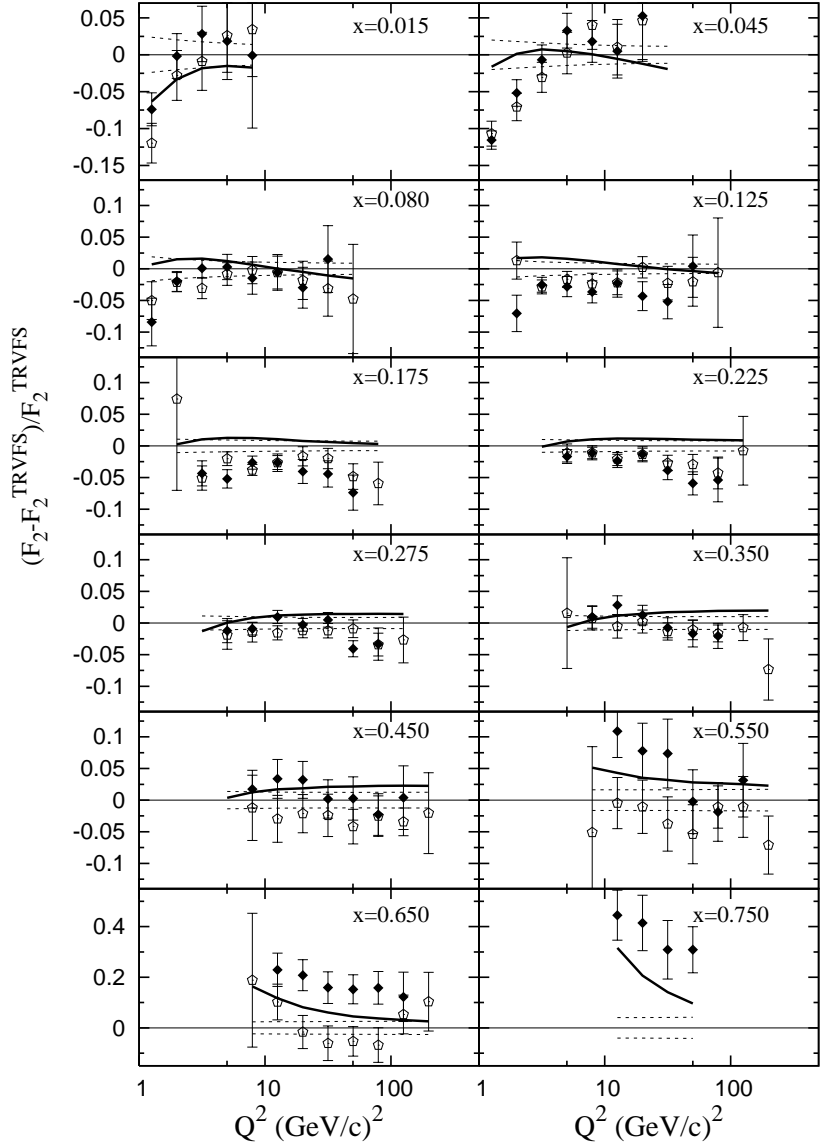,width=0.6\columnwidth
}}
\vspace{0.3cm}
\caption{$F_2(x,Q^2)$ fractional difference
$\frac{F_2-F_2^{TRVFS}}{F_2^{TRVFS}}$ with respect to the
TRVFS(MRST2001E) model. Data points are NuTeV (solid dots) and CCFR
(open circles). Theory curves are ACOTFFS(CTEQ5HQ1) (solid line)
and TRVFS(MRST2001E) $\pm 1\sigma$ (dashed lines).
Theory curves are corrected for target mass and nuclear effects.
}
\label{compf2}
\end{figure}

\begin{figure}
\centerline{
\psfig{figure=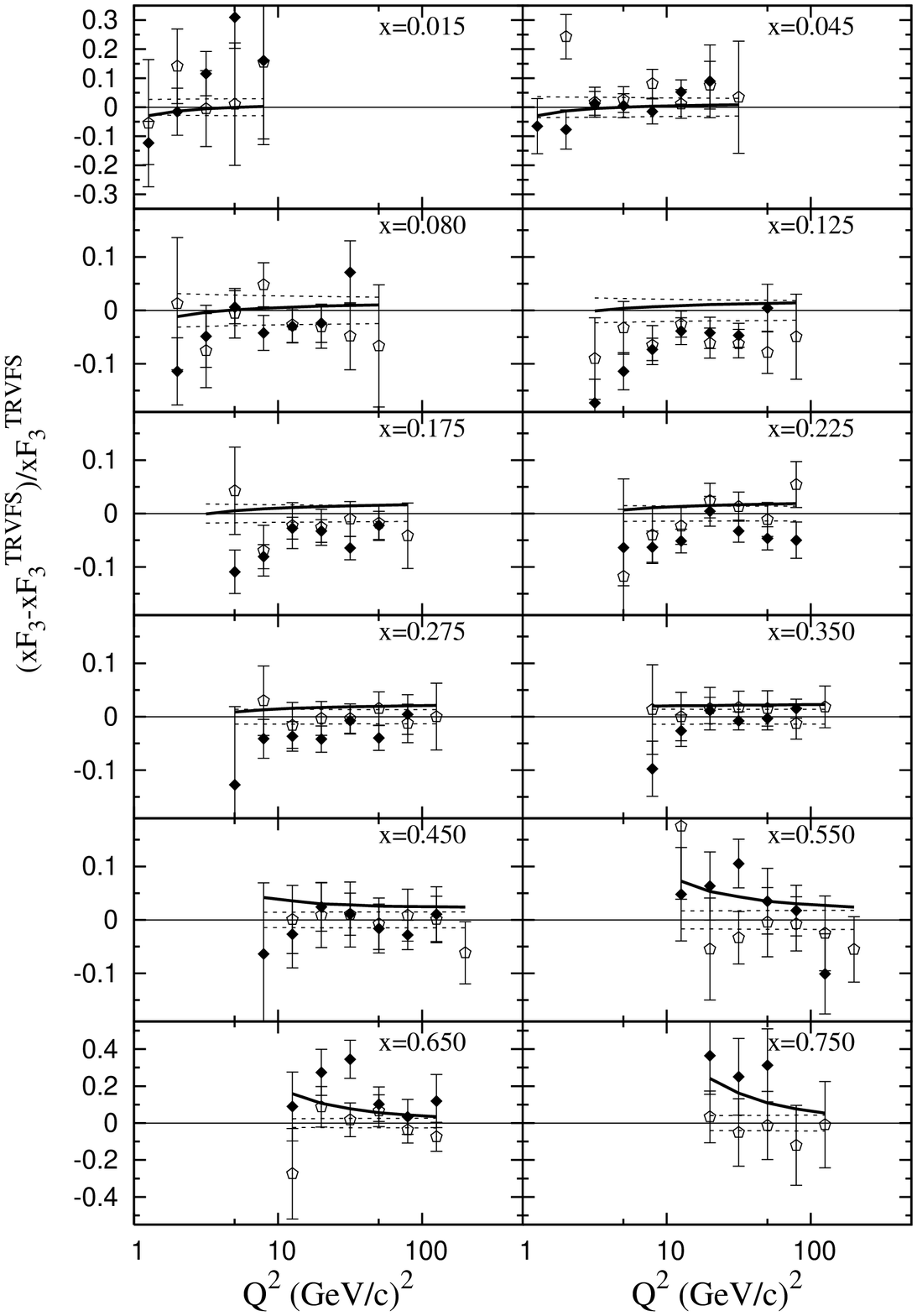,width=0.6\columnwidth
}}
\vspace{0.3cm}
\caption{$xF_3(x,Q^2)$ fractional difference
$\frac{xF_3-xF_3^{TRVFS}}{xF_3^{TRVFS}}$ with respect to the
TRVFS(MRST2001E) model. Data points are NuTeV (solid dots) and CCFR
(open circles). Theory curves are ACOTFFS(CTEQ5HQ1) (solid line)
and TRVFS(MRST2001E) $\pm 1\sigma$ (dashed lines).
Theory curves are corrected for target mass and nuclear effects.
}
\label{compxf3}
\end{figure}

Figures \ref{compf2} and \ref{compxf3} show a comparison of NuTeV and CCFR data
with two theory models. The plot on Figure \ref{compf2} shows the ratio of $F_2(x,Q^2)$
to the Thorne-Roberts variable-flavor scheme (TRVFS)
NLO QCD model \cite{trvfs} using the MRST2001 NLO 
$PDF$ set \cite{mrst2001e} for all $x$ bins.
The $1\sigma$ error band from the $PDF$ set is also shown. 
The second theory curve shown on the plot is ACOT fixed-flavor scheme 
(ACOTFFS) 
NLO model \cite{acot} using CTEQ5HQ1 
\cite{cteq5} $PDF$s. Similarily, the plot on Figure \ref{compxf3} shows the
same comparison for $xF_3(x,Q^2)$.
All theory models have been corrected for target-mass effects using 
the Georgi-Politzer model \cite{gp_tm}. These are 
important at high-$x$ and low-$Q^2$. For example, they increase the
theory prediction by about 5\% at $x=0.65$ and $Q^2=15$~GeV$^2$.
To make a direct comparison,
the theory curves are also corrected for nuclear target effects.
A standard treatment of nuclear effects, which are not well determined
for neutrino scattering, is to apply corrections 
measured in charged-leptons scattering 
from nuclear targets \cite{nurev}.
We use a multiplicative correction factor
of the form \cite{seligthesis}
\begin{equation}
f(x)=1.10 - 0.36x - 0.28e^{-21.9x}+2.77x^{14.4}
\label{eqn:nucl}
\end{equation}
obtained from a fit to charged-lepton scattering on nuclear targets.
The correction is independent of $Q^2$ and is 
small at intermediate $x$ but is large at
low and high $x$, (10\% at $x=0.015$ and 
increases from 7\% at $x=0.45$ to 15\% at $x=0.65$).

The data show reasonable agreement with both the TRVFS(MRST2001E) and 
the ACOTFFS(CTEQ5HQ1)
NLO QCD calculations for most of the $x$ and $Q^2$ range.
At low $x$ ($x=0.015$ and $x=0.045$) both NuTeV and CCFR results have different $Q^2$ dependence than theoretical predictions.
At high-$x$ ($x>0.5$) our data are systematically
above the theory predictions. Compared with the  TRVFS(MRST2001E) model
the data are 15-20\% high at $x=0.65$ and 
25-40\% at $x=0.75$.
The data are about 10\% higher at $x=0.65$ and 15\% at $x=0.75$ than 
the ACOTFFS(CTEQ5HQ1) model prediction.
The  $Q^2$ dependence of the high-x data is similar to the prediction 
from the ACOTFFS(CTEQ5HQ1) model.
A different theoretical treatment of nuclear effects could make a 
sizable difference at small and large $x$. 
NuTeV perhaps indicates that neutrino scattering 
favors smaller nuclear effects at high-$x$
than are found in charged-lepton scattering. 
At small $x$, new theoretical calculations show that in the shadowing 
region the nuclear correction has $Q^2$ dependence \cite{nuclear2,nuclear3}.
The standard nuclear correction obtained from a fit to charged lepton data implies a suppression of 10\% independent of $Q^2$ at $x=0.015$, while 
for $x=0.015$ reference \cite{nuclear3} finds a suppression of 15\% at
$Q^2=1.25$GeV$^2$ and suppression of 3.4\% at $Q^2=7.94$GeV$^2$.
This effect somewhat improves agreement with data at low-$x$.

\section{Data Access}

The NuTeV neutrino and antineutrino 
cross sections and point-to-point covariance matrix can be 
downloaded from reference \cite{nutevresults}.
The tar file contains an unpacking routine and information 
on how to use the results.

\newpage
\appendix
\section*{NuTeV Cross Section Model}

The NuTeV cross section model is inspired by
the LO parameterization prescribed by Buras and Gaemers in reference \cite{bg}. The leading order model is modified to include non-leading order
effects from $R_L$, higher-twist contributions, and the charm mass
as described below. Because the quark densities used in the model
are obtained from fits to neutrino-iron scattering data, no external
model for nuclear effects is required to describe the NuTeV data.
The cross section model described here is the `Born'-level 
neutrino DIS cross section for an isoscalar target. In addition,
we correct for radiative effects using reference \cite{bardin}.

The neutrino isoscalar structure functions 
are given by
\begin{eqnarray}
2xF_1(x,Q^2)& = & xu_v(x,Q^2) +  xd_v(x,Q^2) \nonumber \\
&+&  2[xu_s(x,Q^2) +  xd_s(x,Q^2) + xs_s(x,Q^2) ], \nonumber \\
F_2(x,Q^2)&=& 2xF_1(x,Q^2) \times  \frac{1+R_L(x,Q^2)}{1+4M^2x^2/Q^2}, \nonumber \\
xF_3 (x,Q^2)& = &  xu_v(x,Q^2) +  xd_v(x,Q^2). \nonumber
\end{eqnarray}
\noindent
The CKM matrix elements are used in the above formula to account for the mixing between the quarks, even though they are not shown here.
$R_L(x,Q^2)$ is obtained from 
an empirical fit to world data as described in reference \cite{rworld}
\begin{eqnarray}
R_{L}(x,Q^2) &=& \frac{0.0635}{{\rm ln}(Q^2/0.04)}\Theta(x,Q^2) \nonumber \\
&+& \frac{0.5747}{Q^2} - \frac{0.3534}{Q^4+0.09}, \nonumber \\
\Theta(x,Q^2)&=&1+12(\frac{Q^2}{1+Q^2})(\frac{0.125^2}{x^2+0.125^2}). \nonumber
\end{eqnarray}
\noindent

The charm production cross section is calculated using the slow
rescaling model \cite{nurev}. The Bjorken scaling variable, $x$, is rescaled
\begin{eqnarray}
 x\rightarrow \xi=x \cdot (1+\frac{m_c^2}{Q^2}). \nonumber
\end{eqnarray}
The charm production differential cross section is 
suppressed by the factor $1-\frac{m_c^2}{2ME\xi}$.
The value of the charm mass parameter used, $m_c=1.40\pm 0.18$,
is obtained from the weighted average of 
leading-order experimental measurements \cite{rab},\cite{locharm}. 

To account for higher-twist effects at high $x$ and low $Q^2$ an
 empirical model is used to constrain the $Q^2$ dependence of the $PDF$s
by rescaling $x$ following reference \cite{arie}.
\begin{eqnarray}
x\rightarrow \xi_{HT}=x\frac{Q^2+B_{HT}}{Q^2+A_{HT}\cdot x},\nonumber
\end{eqnarray}
where $A_{HT}$ and $B_{HT}$ are fit parameters in the model.
The shape of GRV94L0 \cite{grv94lo} $PDF$s 
is used to extrapolate the model 
for $Q^2<1.35 GeV^2/c^2$.  The normalizations
are constrained by matching the $PDF$s to the Buras-Gaemers 
$PDF$s at $Q^2=1.35 GeV^2/c^2$.

The form for the parton distribution functions follow
the Buras-Gaemers parameterization in reference \cite{bg}.
The valence distributions are given by
\begin{eqnarray}
x u_v(x)& = & u_{tot} \times [ x^{E_1}(1-x)^{E_2} + AV_2x^{E_3}(1-x)^{E_4}] \nonumber \\
x d_v(x)& = & d_{tot} \times x u_v(x) \times (1-x), \nonumber
\end{eqnarray}
with parameters $AV_2$, $E_1$, $E_2$, $E_3$, and $E_4$. 
The $Q^2$-dependence for the valence is given by
\begin{eqnarray}
E_i&=&E_{i0}+E_{i1} \cdot s, i=1,4 \nonumber \\
s &=& {\rm ln}\left[ \frac{{\rm ln}(\frac{Q^2}{A_0^2})}{{\rm ln}(\frac{Q_0^2}{A_0^2})}\right] \nonumber
\end{eqnarray} 
where the reference $Q_{0}^{2}$ is set 
to 12.6 $GeV^{2}/c^{2}$ and the QCD scale, $A_{0}$, is
a parameter in the fit.

The constraints on the valence distributions include the 
normalization of the valence densities, which is determined from a variant
of the Gross-Llewllyn-Smith sum rule, and the relative normalization between
$u_v$ and $d_v$, which is determined from quark counting

\begin{eqnarray}
\int_0^1 xF_3(x,Q^2) dx&=&3\Big(1-\frac{\alpha_s(Q^2)}{\pi}\Big) \nonumber \\
   &=& 3 \Big( 1 - \frac{A1}{{\rm log} (Q^2/A_0^2)}
      - \frac{A2}{ \left[ {\rm log} (Q^2/A_0^2) \right] ^2 } \Big) \nonumber \\
\int_0^1  u_v(x,Q^2) dx &=&
        2 \Big( 1 - \frac{A1}{{\rm log} (Q^2/A_0^2)}
      - \frac{A2}{ \left[ {\rm log} (Q^2/A_0^2) \right] ^2 } \Big) \nonumber \\
\int_0^1 d_v(x,Q^2) dx 
       & = & \Big( 1 - \frac{A1}{{\rm log} (Q^2/A_0^2)}
      - \frac{A2}{ \left[ {\rm log} (Q^2/A_0^2) \right] ^2 } \Big)\nonumber
\end{eqnarray}
\noindent
In addition a charge constraint is required
\begin{eqnarray}
\frac{2}{3}\int_0^1  u_v(x,Q^2) dx &-& \frac{1}{3}\int_0^1 d_v(x,Q^2) dx \nonumber \\
         &=& \Big( 1 - \frac{A1}{{\rm log} (Q^2/A_0^2)}
      - \frac{A2}{ \left[ {\rm log} (Q^2/A_0^2) \right] ^2 } \Big),\nonumber
\end{eqnarray}
\noindent
where the fit parameters $A1$ and $A2$ describe 
the normalization of the leading order
and next-to-leading order
terms respectively.

The light quark sea distributions are given by
\begin{eqnarray}
x\overline u(x) &=& x\overline d(x)  =  \frac{1}{2(\kappa+2)} xS(x) \nonumber \\
      & = &  \frac{1}{2(\kappa+2)} (AS (1-x)^{ES} + AS_2 (1-x)^{ES_2}), \nonumber
\end{eqnarray}
\noindent
where $S(x)$ is the light-quark sea density. 
$AS$, $AS_2$, $ES$, and $ES_2$ are defined in terms of
fit parameters below, and the parameter $\kappa$
is the relative normalization to the strange sea, also defined below.
\noindent
The light quark sea distributions
can be determined from the first two moments, because they decrease rapidly with x
\begin{eqnarray}
S_2&=&\int_0^1 x S dx=\frac{3}{4}D_{22}+\frac{1}{4}D_{12}, \nonumber \\
S_3&=&\int_0^1 x^2 S dx=\frac{3}{4}D_{23}+\frac{1}{4}D_{13}, \nonumber 
\end{eqnarray}
where the values for $D_{ij}$, which are completely specified by LO QCD, are given in \cite{bg}. $AS_2$ and $ES_2$ are evolved with $Q^2$ in the following way
\begin{eqnarray}
AS_2&=&AS_{20}+AS_{21}{\rm ln}(Q^2), \nonumber \\
ES_2&=&ES_{20}+ES_{21}{\rm ln}(Q^2). \nonumber
\end{eqnarray}
$S_2$, $S_3$, $AS_{20}$, $AS_{21}$, $ES_{20}$, and $ES_{21}$ are parameters in the fit.
$AS$ and $ES$ are constrained to match the moments $S_2$ and $S_3$.
\begin{eqnarray}
ES&=& \frac{S_2-AS_2/(ES_2+1)}{S_3-AS2/((ES_2+1)(ES_2+2))}-2,  \nonumber \\
AS&=&(ES+1)\frac{S_2-AS_2}{ES_2+1}. \nonumber
\end{eqnarray}

The strange sea distribution can be measured from the dimuon inclusive cross
section. The parameterization of the strange sea is a LO fit to the CCFR dimuon differential cross section \cite{rab},
\begin{eqnarray}
x s(x) &=& x \overline s(x) =  \frac{\kappa}{2(\kappa+2)} xSS(x) \nonumber \\
\end{eqnarray}
where $SS(x)$ is the strange sea density.
The relative normalization to the light quark seas is determined by the
parameter $\kappa$,
\begin{eqnarray}
xs(x) \propto   \kappa \frac{x \overline u_R(x)+x \overline d_R(x)}{2}(1-x)^\alpha, 
  \nonumber
\end{eqnarray}
where $\alpha$ describes the shape of the strange sea.
Here the light quark sea densities 
$\overline u_R(x)$ and $\overline d_R(x)$, and the values of $\alpha=2.5$ and $\kappa=0.373$ are obtained from reference \cite{rab}.

Momentum sum rule is constrained by the second moment of the gluon distribution
$G_2=\int_0^1 xg(x,Q^2)dx$

\begin{eqnarray}
G_2&+&\int_0^1 \frac{1+R(x,Q^2)}{1+4M^2x^2/Q^2} [xu_v(x,Q^2)\nonumber \\
    &+&xd_v(x,Q^2)+xS(x,Q^2)]dx=1  \nonumber
\end{eqnarray}

Additional small corrections are used to modify $u$ and $d$ distributions to account for the $u,d$ asymmetry observed in muon DIS and Drell-Yan data. Drell-Yan data from E866 \cite{e866} is used to constrain the $\overline{u}, \overline{d}$ asymmetry by the factor
\begin{eqnarray}
f(\overline{d}/\overline{u})&=&\frac{1}{{\rm max}(1-x(2.7-0.14 {\rm ln} (Q^2)-1.9x),0.1)} \nonumber
\end{eqnarray}
The modified light sea distributions $\overline{u'},\overline{d'}$ are constrained by $\overline{u'}+\overline{d'}=\overline{u}+\overline{d}$ and have the form
\begin{eqnarray}
\overline{u'}&=&\overline{u}\left(\frac{\overline{u}+\overline{d}} {\overline{u}+\overline{d} \cdot f(\overline{d}/\overline{u})}\right), \nonumber \\
\overline{d'}&=&\overline{d}\left(\frac{\overline{u}+\overline{d}} {\overline{u}+\overline{d} \cdot f(\overline{d}/\overline{u})}\right)\cdot
 f(\overline{d}/\overline{u}). \nonumber
\end{eqnarray}
The valence distributions are also modified for the $u_v,d_v$ asymmetry. The modified valence distributions $u_v',d_v'$ are constrained by $u_v'+d_v'=u_v+d_v$ and have the form \cite{ccfrxsec}
\begin{eqnarray}
u_v'&=&\frac{u_v}{1+\delta(d/u) \cdot u_v/(u_v+d_v)}, \nonumber \\
d_v'&=&\frac{d_v+u_v \cdot \delta(d/u)}{1+\delta(d/u) \cdot u_v/(u_v+d_v)}, \nonumber \\
\delta(d/u)&=&0.12079-1.3303x+4.9829x^2 \nonumber \\
           &-& 8.4465x^3+5.7324x^4 \nonumber.
\end{eqnarray}

The model fit parameters are obtained after an iterative loop in which 
the $n^{th}$ loop flux and cross section are re-extracted 
with the $(n-1)^{th}$ loop model fit parameters. The $n^{th}$ loop model then 
determines new acceptance and smearing corrections which are used 
for the $(n+1)^{th}$ cross section and fit. 
New radiative corrections are computed to correspond to each new model.
The initial flux and cross section are determined
using a starting model with the Buras-Gaemers $PDF$ parameters from 
the best fit to CCFR data \cite{ccfrxsec}.
The process is iterated until the average relative change in the cross
section data points compared to the previous iteration is less than
0.1\% (within 3 iterations). 

To constrain the high-$x$ and low $Q^2$ part of the model, which 
is important in order to model our flux data sample, we include data from 
charged-lepton scattering in the fit. The SLAC \cite{slac}, BCDMS \cite{bcdms}, and
NMC \cite{nmc} overlap with the kinematic range of our
data set but extend to lower-$Q^2$.
The charged-lepton data for $x$ in the range $0.4<x<0.7$,
are included in the fit $\chi^2$ function along with the NuTeV data
set to determine the best fit parameters. The charged-lepton data
must first be corrected to $F_2^\nu$ using our model
and corrected to an $A=56$ iron target using Equation \ref{eqn:nucl}.
The normalization of the charged-lepton data
relative to the NuTeV data is unconstrained in the fit.

Table \ref{tab:bgpars} gives the NuTeV cross section model
fit values for each parameter and their estimated uncertainties.
A total of 19 parameters are fit in the model.
Note that we do not expect that the parameterization 
extrapolated beyond the region of the NuTeV data set
will be a good description of data there.

\begin{table}[b]
\caption{The final fit parameters for the NuTeV cross section model.}
\label{tab:bgpars}
\begin{center}
\begin{tabular}{ c c c }
{\bf Parameter} & {\bf Value} & {\bf Estimated Error} \\ \hline
$A_0$   &  0.583  &  0.017 \\
$A_1$   &  0.295   &  0.013 \\
$A_2$   &  0.17   &  0.03 \\
$E_{10}$ &  0.5333 &  0.0025 \\
$E_{11}$ &  -0.028 &  0.011 \\
$E_{20}$ &  2.61 &  0.015 \\
$E_{21}$ &  1.31  &  0.045 \\
$AV_2$  & 637.0  &  75.0 \\
$E_{30}$ &   4.56 &  0.14 \\
$E_{40}$ &  12.5  &  0.35 \\
$S_2$   &  0.1625  &  0.0013 \\
$S_3$   &   0.0159 &  0.0004 \\
$G_3$   &  0.031 &  0.003 \\
$AS_{20}$ &  1.06 &  0.11 \\
$AS_{21}$ &  1.76 &  0.25 \\
$ES_{20}$ &  185.0  &  20.0 \\
$ES_{21}$ &  8.4  &  8.0 \\
$A_{HT}$ &   1.187  &  0.035 \\
$B_{HT}$ &  0.33 &  0.02\\ 
\end{tabular}
\end{center}
\end{table}

\end{document}